# A 4565 Myr old andesite from an extinct chondritic protoplanet.

Jean-Alix Barrat[a*], Marc Chaussidon[b], Akira Yamaguchi[c], Pierre Beck[d], Johan Villeneuve[e], David J. Byrne[e], Michael W. Broadley[e], Bernard Marty[e]

[a] Univ Brest, Institut Universitaire Européen de la Mer (IUEM), UMR 6539, Place Nicolas Copernic, F-29280 Plouzané, France ;[b] Université de Paris, Institut de physique du globe de Paris, CNRS, F-75005 Paris;[c] National Institute of Polar Research, Tokyo, 190-8518, Japan ;[d]Universite Grenoble Alpes, CNRS, Institut de Planetologie et d'Astrophysique de Grenoble (IPAG), Saint-Martin d'Heres, France;[e] Université de Lorraine, CNRS, CRPG, F-54000 Nancy, France.

*Corresponding author Jean-Alix Barrat

**Email:** barrat@univ-brest.fr

**Abstract**

The age of iron meteorites implies that accretion of protoplanets began during the first millions of years of the solar system. Due to the heat generated by $^{26}$Al decay, many early protoplanets were fully differentiated, with an igneous crust produced during the cooling of a magma ocean, and the segregation at depth of a metallic core. The formation and nature of the primordial crust generated during the early stages of melting is poorly understood, due in part to the scarcity of available samples. The newly discovered meteorite Erg Chech 002 (EC 002) originates from one such primitive igneous crust, and has an andesite bulk composition. It derives from the partial melting of a non-carbonaceous chondritic reservoir, with no depletion in alkalis relative to the sun's photosphere, and at a high melting rate of around 25%. Moreover, EC 002 is to date, the oldest known piece of an igneous crust with a $^{26}$Al-$^{26}$Mg crystallization age of 4565.0 Myr. Partial melting took place at 1220 °C, up to several hundred kyr before, implying an accretion of the EC 002 parent body circa 4566 Myr ago. Protoplanets covered by andesitic crusts were probably frequent. However, no asteroid shares the spectral features of EC 002, indicating that almost all of these bodies have disappeared, either because they went on to form the building blocks of larger bodies or planets, or were simply destroyed.

**Significance Statement**

The crusts of the oldest protoplanets are virtually unknown due to the scarcity of samples. Here we describe the oldest known lava that crystallized ca. 4565 Ma ago, and formed by partial melting of a chondritic parent body. $^{26}$Al-$^{26}$Mg systematics suggest that the elapsed time between melting and crystallization was significant, on the order of several $10^5$ yr, probably due to the viscosity of the magma. Although the first protoplanetary crusts were frequently not basaltic, their remains are not detected in the asteroid belt, because their parent bodies served as the building-blocks for larger rocky bodies, or were nearly totally destroyed.

**Main Text**

**Introduction**

Despite the large number of samples in the meteorite record that originate from the crust or mantle of rocky bodies (about 3100 are known today), these rocks provide an incomplete picture of the diversity of the differentiated bodies that formed in the early solar system (1). Indeed, about 95% of these meteorites originate from only two bodies, with 75 % coming from the crust of a single asteroid (possibly 4-Vesta), and the other 20% from the mantle of a presumably large object, the now-destroyed ureilite parent body (2, 3). Thus, until recently, known achondritic lavas were essentially basalts (eucrites) from 4-Vesta, and a handful of other basaltic rocks from unknown parent bodies (the angrites and some ungrouped achondrites such as Northwest Africa 011 (4) or Ibitira (5). Although certainly not representative of the magmatic activity of all the planetesimals, these achondritic lavas strengthened the general view that their crusts were essentially basaltic in composition. However, the discovery of some rare achondrites of andesitic or trachyandesitic composition (e.g., Graves Nunataks 06128 and 016129 (6, 7), ALM-A (8), Northwest Africa (NWA) 11119 (9)), demonstrated that the diversity of the lavas formed on protoplanets may have been more important than previously thought. Experimental studies motivated by these new meteorites have shown that the generation of silica-rich liquids is possible from the melting of chondrites (10-12). Thus, the formation of andesitic crust was possibly common on protoplanets, especially for those that were not Na and K depleted (12), contrary to what the meteorite record suggests. However, the processes that built such a crust, and the genesis of protoplanetary andesites are not well known due to the rarity of the samples. Here we report on Erg Chech 002 (EC 002), a unique andesite achondrite found in the spring of 2020 in the Sahara. This meteorite is the oldest magmatic rock analyzed to date, and sheds light on the formation of the primordial crusts that covered the oldest protoplanets.

**Results**

**Petrography.** EC 002 is an unbrecciated greenish rock with a medium-grained groundmass (grain size in the order of 1-1.5 mm) with 8 vol% pores. Its texture suggests a crystallization within a thick flow or in a shallow intrusion (Fig. 1). It consists of 45 vol% of lath-shaped albitic feldspar (plagioclase and possibly anorthoclase) containing lamellae of K-feldspar, 38 vol% of anhedral pyroxene (ca. 0.6 x 3.7 mm), 5 vol% interstitial silica minerals (cristobalite and tridymite (PO) identified by Raman spectroscopy (SI appendix, Fig. S1 and S2), with no quartz), and minor spinel, ilmenite, Ca-phosphate, troilite and FeNi metal. The samples we have examined are rather fresh, with only a few rust patches produced by the alteration of troilite and metal, and some carbonate fillings in the fractures, which are typical of Saharan finds (13). Pyroxenes are partly equilibrated and consist of relict augite with fine, closely spaced (~1 µm thick) exsolution lamellae of low-Ca pyroxene, or relict low-Ca pyroxene with fine exsolution lamellae of augite. In addition, EC 002 also contains angular to rounded pyroxene and olivine xenocrysts, irregularly dispersed in the groundmass (Fig. 1 and 2), and reaching several centimeters (up to 9 cm (14)). They are generally mantled by fine-grained groundmass pyroxenes. EC 002 is slightly shocked. Plagioclase and pyroxene show only moderate mottled extinction. Some plagioclase grains have fine polysynthetic twinning. Fractures in these minerals are not prominent. The shock stage is M-S2 (15).

**Phase compositions** (SI Appendix, Fig. S3-S7, Table S1-S4). Bulk groundmass pyroxene compositions are generally augitic ($Wo_{27.7-33.0}En_{36.9-38.5}Fs_{30.1-31.0}$). Some crystals exhibit remnant zoning from core to rim ($Wo_{22.9}En_{39.4}Fs_{37.7}$ to $Wo_{35.2}En_{36.3}Fs_{28.5}$). The crystals are partly equilibrated and are a mixture of augite ($Wo_{38.7-40.9}En_{39.5-41.5}Fs_{18.8-20.7}$) and fine exsolutions of low-Ca pyroxene ($Wo_{2.4-4.3}En_{47.1-49.4}Fs_{47.3-49.3}$). Feldspars are albitic ($Or_{2.0-7.1}Ab_{75.8-87.4}An_{6.7-21.6}$) and contains lamellae of K-rich feldspar ($Or_{84.0-84.4}Ab_{11.3-11.6}An_{4.3-4.40}$). Groundmass spinel compositions are $Usp_{18.2-56.5}Sp_{2.6-6.2}Cm_{94.2-96.5}$ and Mg# = 3.51-5.83, with no Zn (<0.02 wt%). Silica phases contain detectable amounts of Al, Fe, Ca and alkalis ($Al_2O_3$ = 1.8-2.7 wt%; $K_2O$ < 0.5 wt%; $Na_2O$ = 0.7-1.4 wt%; CaO = 0.03-0.15 wt%; FeO < 0.3 wt%).

We analyzed two xenocrysts: fragments of a 1 cm-long olivine megacryst and a small pyroxene. The olivine is forsteritic ($Fo_{88}$, FeO/MnO=24.2 wt%/wt%), Ca rich (CaO = 1.3 wt%), poor in Ni (1.3 µg/g) and Co (10.2 µg/g). Its light rare earth element (REE) and alkali abundances are much higher than expected (La= 11.7 ng/g, K =102 µg/g, Rb= 4.9 µg/g, Cs = 0.28 µg/g) and could reflect the presence of melt inclusions. The small pyroxene is a low-Ca pyroxene (Fig. 2). It displays a chemically homogeneous core ($Wo_{4.7}En_{76.8}$) with zoned rims, mantled by groundmass pyroxenes. The cores have relatively high $Cr_2O_3$ (~0.5 wt%) and $Al_2O_3$ (~0.3 wt%), and low $TiO_2$ contents (~0.1 wt%) compared to low-Ca pyroxene in the groundmass. The FeO/MnO ratios of the xenocryst core (21.1 wt%/wt%) are identical to those of the low-Ca pyroxene in the rims and in groundmass (~20-21 wt%/wt%). Pyroxenes surrounding the crystal are more magnesian than those in the groundmass. Spinel grains found inside this xenocryst are more Cr and Al rich those found in the groundmass ($Usp_{0.9-2.9}Sp_{12.4-17.3}Cm_{84.3-88.1}$ and Mg# =11.9-15.7).

These olivine or pyroxene xenocrysts could be debris from the nascent mantle of the protoplanete (melting residues) or crystals formed from previous magmas. A definitive answer cannot be given with the study of only two crystals. Note that the largest pyroxene xenocrysts reach 9 cm (14), and are certainly not "mantle-like". Hence, xenocrysts are more likely derived from the crystallization of magma(s) more magnesian than EC 002, and deriving from nearby source(s) (similar FeO/MnO ratio and $\Delta^{17}O$ (14)).

**Geochemistry**. We crushed a 1.1 g sample devoid of apparent xenocrysts, representative of the groundmass, and therefore very close to the melt from which EC 002 crystallized. With 58 wt% $SiO_2$ and 4.54 wt% of $Na_2O+K_2O$ (Fig. 3, and SI Appendix, Table S4), it is andesitic according to the International Union of Geological Sciences criteria (18). The rock is quite rich in MgO and FeO. Its CIPW norm indicates proportions of pyroxenes and plagioclase in accordance with modal estimates (53 wt% pyroxene and 40.5 wt% plagioclase, i.e. about 47 vol% pyroxene and 47 vol% plagioclase taking into account the mineral densities). The calculation also shows some normative quartz (2.5 wt%) indicating a rock slightly oversaturated in $SiO_2$. EC 002 is very poor in P (216 µg/g), Ni (18.5 µg/g), Co (5.85 µg/g), Cu (1.4 µg/g), Pb (90 ng/g), W (22 ng/g), Ga (2.6 µg/g), and Zn (0.44 µg/g). EC 002 displays some excesses in Ba, Sr and U, which are usual for Saharan finds (e.g., 19), and will not be further discussed. Other incompatible trace element abundances (e.g., REE, Th...) are low and of the order of 5 or 6 times the chondritic reference. Indeed, the CI-normalized trace element pattern of EC 002 is rather flat with no noticeable anomaly, even for high field strength elements and alkalis (Fig. 4). This rock is only slightly light REE depleted ($La_n/Sm_n$ = 0.94) with a small positive Eu anomaly (Eu/Eu*=1.05). Moreover, it displays the same Tm negative anomaly (Tm/Tm*=0.973) as non-carbonaceous chondrites (Tm/Tm*<1 and typically 0.97-0.98) and achondrites, and inner solar system planetary bodies (Vesta, Mars, Moon, Earth, with average Tm/Tm* typically ≈ 0.975, (21)).

The chemical composition of EC 002 is very different from that of other andesitic achondrites. It is, for example, much richer in alkalis than NWA 11119, without reaching the concentrations of ALM-A or GRA 06 (Fig. 3). The smoothness of its trace element pattern (Fig. 4), parallel to that of non-carbonaceous chondrites even for alkalis, also distinguishes it from the other known achondritic lavas.

**$^{26}$Al-$^{26}$Mg systematics.** Nineteen feldspar grains and eleven pyroxenes were analyzed (SI Appendix, Tables S5-S6). The pyroxenes are characterized by very low $^{27}$Al/$^{24}$Mg ratios ranging from 0.018 to 0.06 and homogeneous (within their typical ±0.15‰ 2 SE errors) $\delta^{26}$Mg* from -0.10 to +0.25‰ (average +0.067±0.076 ‰ (2SE)). Plagioclases display a wide range of radiogenic Mg enrichments, with $^{27}$Al/$^{24}$Mg from 1597 to 5316 and $\delta^{26}$Mg* from 65 to 215 ‰. Pyroxenes and plagioclase define a $^{26}$Al isochron (Fig. 5), with an initial $^{26}$Al/$^{27}$Al ratio of (5.72 ± 0.07) × 10$^{-6}$, the highest value ever reported for an achondrite. This ratio translates into a closure age of the $^{26}$Al-$^{26}$Mg system of 4565.0 Myr (i.e. 2.255 ± 0.013 Myr after Ca-Al-rich inclusions (CAIs)) assuming a canonical distribution of $^{26}$Al (22), or of 4566.1 Myr (i.e. 1 Myr after CAIs) using the D'Orbigny angrite anchor (23, 24). The calculated initial $\delta^{26}$Mg* ($\delta^{26}$Mg*$_0$ = +0.065 ± 0.08‰ (2SE)) is consistent with the average composition of the pyroxenes and significantly higher than the canonic initial of the solar system of -0.040 ± 0.029‰ as defined from CAIs (22).

**Noble gas isotope composition.** Abundances and isotope ratios of He, Ne and Ar were analyzed by step heating (SI Appendix, Table S7). Isotope ratios of He and Ne are strongly enriched in cosmogenic isotopes, with concordant $^{3}$He and $^{21}$Ne exposure ages of 26.0 ± 1.6 Myr and 25.6 ± 1.0 Myr respectively. The bulk $^{40}$Ar/$^{36}$Ar ratio (5522 ± 42) is strongly enriched in radiogenic $^{40}$Ar from the decay of $^{40}$K. The $^{40}$Ar abundance corresponds to a K-Ar age of $4534^{+117}_{-125}$, within uncertainty of the closure age for the $^{26}$Al-$^{26}$Mg system. This indicates that Ar has been effectively retained within EC 002 since shortly after formation, precluding significant thermal events and implying a relatively early breakup of the parent body. Furthermore, the absence of significant trapped 'planetary' noble gas components suggests that the parent melt of EC 002 was efficiently degassed prior to crystallization. After correction for cosmogenic $^{36}$Ar, trapped $^{36}$Ar abundance is only 4.3 x10$^{-13}$ mol.g$^{-1}$, which is more than an order of magnitude lower than typical achondrite abundances (25).

**Spectroscopy** (SI Appendix, section B). Reflectance spectra of EC 002 reveal the presence of two strong absorptions related to Ca-rich pyroxene. This spectral signature does not correspond to any known asteroid spectral type. Comparison between colors of EC 002 and those from about 10000 objects from the SDSS database also reveal the rarity if not absence of similar objects within the asteroid population.

**Discussion**

As demonstrated by its chemistry and Al-Mg age, EC 002 is a unique fragment of the crust of an ancient differentiated body, contemporaneous with the formation of the cores of the parent bodies of iron meteorites (26). The oldest igneous rock previously described was NWA 11119 (9), with an age 1.24 Myr younger than EC 002 when calculated with the same canonical compositions. For major elements, EC 002 is chemically distinct from all the other basaltic, andesitic or trachyandesitic achondrites (Fig. 3). Its alkali to refractory incompatible element ratios are unfractionated making EC 002 very different from the crust of telluric planets (Fig. 4) in the sense that it is not depleted in volatile elements. For example, its K/Th ratio (=22500) is similar to that of CI chondrites (=19400, (20)) and much higher than that of the crusts of Mercury, Earth or Mars (≈ 5500 or below), as well as the Moon, Vesta and the parent body of the angrites, which are strongly alkali-depleted (27, 28). However, its very low Zn and Ga abundances point to a possible depletion for some of the other volatile elements, inherited from its accretionary materials. These low concentrations could also be the fingerprints of the early segregation of some metal and sulfides (initiation of the core formation), thus explaining the low abundances of Ni, Co and W. EC 002 displays a Tm/Tm* ratio identical to those of differentiated bodies from the inner solar system, and clearly points to a non-carbonaceous affinity (21). Thus, despite Ti, Cr or Mo isotopic data not yet being available, the parent body of EC 002 can most likely be affiliated to the non-carbonaceous (NC) family of objects formed early in the accretion disk (e.g., 29; 30). Oxygen isotopic composition of EC 002 is consistent with this interpretation. It is similar to that of some ungrouped basaltic (NC) achondrites such as Bunburra Rockhole, Asuka 881394 and Emmaville, but which are, however, mineralogically and chemically too different to originate from the same parent body as EC 002 (12).

The genesis of EC 002 allows us to understand some aspects of primordial crust formation on bodies of chondritic compositions. Though an andesite can derive from the fractional crystallization of a more primitive melt, this cannot be the case for EC 002. Firstly, EC 002 is poor in highly incompatible elements (e.g., REE or Th). If it was an evolved lava, its parental melt would have even lower concentrations, which would be difficult to envisage even for a primary basalt. The concentration of MgO (7.06 wt%) and the Mg#-number (=100 Mg/(Mg + Fe), atomic = 52.9) of EC 002 are quite high for an andesite, and do not allow to imagine that important amounts of ultramafic cumulates were extracted. Furthermore, fractional crystallization would have inevitably left its fingerprint in the distribution of trace elements, such as a marked negative Eu anomaly generated by plagioclase crystallization, at variance with the composition of EC 002 (Fig. 4). Therefore, it must be considered that EC 002 could be a primitive or even a primary melt. Partial melting of chondrites at low pressure is perfectly capable of generating andesitic magmas rich in $SiO_2$ and alkalis (10-12). However, the major element composition of EC 002 lies at the high-melting end of the experimental melting trends defined for ordinary H or LL chondrites (Fig. 3 and SI appendix Fig. S8). These trends indicate that melts with the same composition as EC 002 are obtained after plagioclase exhaustion, for high degrees of melting (F) of around 25% (SI Appendix Fig. S9). Such high-melting rates would also explain the unfractionated trace element pattern of EC 002, since all phases with high crystal-melt partition coefficients (such as phosphates or plagioclase) would be exhausted in the source after $\approx$ 17 % partial melting (12, 31). Indeed, the trace element pattern of EC 002 fits with an enrichment by a factor $\approx 1/F$ of a typical NC chondrite composition (Fig. 4), in perfect agreement with the experimental data.

The thermal history of EC 002 appears quite straightforward. Experimental data allow a precise evaluation of the temperature at which the magma formed (SI Appendix, Fig. S10). The MgO of EC 002 gives an estimated melting temperature of 1224 ± 20°C. This high temperature is confirmed by the crystallization temperatures estimated from the bulk compositions of Ca-rich pyroxene (32) in the range of 1149-1229 °C (average = 1186 °C (standard deviation =25°C), slightly lower due to post-crystallization equilibration during cooling). The final equilibration temperature, about 957 °C, is estimated from the highest Ca concentration in pyroxene (32). The cooling of the rock was fast enough to preserve remnant zoning in groundmass pyroxene and the compositions of the core of the small xenocrysts. As a demonstration, we tentatively modeled the zoning profile of Mg# across the small xenocryst exposed in our section (Fig. 2, and SI appendix). The cooling rate is estimated to be about 5°C/yr between 1200 and 1000°C, a value consistent with a thick lava flow or a shallow intrusion. Important additional constraints are brought by silica polymorphs (SI Appendix). Only cristobalite and tridymite (PO) were detected in EC 002, and quartz is totally lacking. As experimentally shown for eucrites (33, 34), cristobalite crystallized at high temperatures. Subsequently, it partially transformed to tridymite above ~900 °C. Since cristobalite easily transforms to quartz (34), the lack of quartz indicates a very fast cooling rate below 900 °C (>0.1-1 °C/day), consistent with the absence of monoclinic tridymite and the possible occurrence of anorthoclase. The most likely explanation for this change in cooling rate is an impact that would have excavated, or more likely ejected, the rock from its parent body, in agreement with the evidence for shock metamorphism (M-S2). The similarity between the K-Ar retention age and the formation age as calculated from Al-Mg, indicates that EC 002 was not significantly heated following formation, further supporting the idea that any ejection event occurred rapidly after formation. The cooling history of this meteorite therefore appears to be short, since it would have cooled for only a few decades before it was probably ejected.

The Mg isotopic composition of EC 002 gives further clues on the timing of differentiation of its parent body. Because of the fast cooling inferred for EC 002, the old $^{26}$Al age of 2.255±0.013 Myr after CAIs can be considered to date the crystallization of the parent melt. This age is tightly constrained from (i) the quality of the $^{26}$Al isochron (Fig. 5a) that also implies a lack of significant $^{26}$Mg redistribution by metamorphism after this fast cooling, and (ii) the fact that EC 002 is related to inner solar system NC bodies which are considered to have formed with a canonic level of $^{26}$Al even in the hypothesis of a heterogeneous distribution of $^{26}$Al in the accretion disk (35). The fact that the $^{26}$Al isochron intercept gives a $\delta^{26}$Mg*$_0$ significantly higher than the canonic initial of the solar system (Fig. 5b) is likely an indication for a protracted history of the parental melt of EC 002 prior to crystallization. In fact, such excesses of $^{26}$Mg have not been observed for bulk chondrites (e.g., 36, 37) but are known in two ancient achondrites (NWA 7325: $\delta^{26}$Mg*$_0$=0.093 ± 0.004‰ (38); Asuka 881394: $\delta^{26}$Mg*$_0$=0.070 ± 0.052‰, 39). They could result from metamorphic perturbations or redistributions (39), or correspond to the protolith isotopic composition in the case of remelting of a crustal reservoir (38, 40). Here, these explanations are not satisfactory, due to the short cooling duration of EC 002, and the evidence that the andesitic magma formed directly from a chondritic source. In addition, because the olivine and pyroxene xenocrysts present in EC 002 have very low Al/Mg ratio (and thus cannot develop significant radiogenic $^{26}$Mg excesses), their eventual assimilation by the EC 002 parental melt is also unable to explain the positive $\delta^{26}$Mg*$_0$. Alternatively, since EC 002 has a super-chondritic Al/Mg ratio, a high $\delta^{26}$Mg*$_0$ value can simply be acquired if the time elapsed between partial melting and crystallization is long enough for $^{26}$Mg radiogenic excesses to develop in the melt. In order to investigate this possibility further, the Mg isotopic evolutionary curves of EC 002 and the Solar System (equivalent to the evolution of regular chondrites) were compared (Fig. 5b). The two curves intersect between 0.95 Ma and 2.25 Ma (the age of crystallization) after the formation of CAIs. Whilst this range is wide due to the error the $\delta^{26}$Mg*$_0$ value of EC 002, it shows that the transfer of magmas between the melting zones and the surface could have been slow, a duration exceeding a few $10^5$ yr being probable. This result confirms and strengthens the inferences made by Collinet and Grove (12), who calculated that the velocity extraction of silica- and alkali-rich magmas in a protoplanet could have been at least 3 orders of magnitude lower than that of basalts (2.3-230 m/Myr vs. 1.1-1100 km/Myr, respectively). Thus, these magmas would have moved slowly, probably only over narrow distances during one half-life of $^{26}$Al, rendering the thickening of a primordial crust by the stacking of flows difficult.

The parent body of EC 002 was certainly not unusual, and despite the scarcity of andesitic achondrites identified to date, it is reasonable to assume that many similar chondritic bodies accreted at the same time, and were capped by the same type of primordial crust (8, 12). However, this andesitic crust might have been only temporary. Thermal models of asteroids which undergo partial melting due to the heat generated by $^{26}$Al decay (e.g. 26, 41) show that if accretion starts before 1 Myr after CAIs, the average temperature of the body reaches at least 1500 °C at 2.2 Myr and can continue to rise after that. This is true whether accretion takes place instantaneously by gravitational instability or at a slower pace by a combination of gravitational instability and pebble accretion. Such temperatures are incompatible with the thermal history reconstructed for EC 002. One solution to this problem would be to consider (i) that the parent body of EC 002 accreted later than 1 Myr (i.e. around 1.5 Myr), and (ii) that EC 002 formed at the surface or close to the surface of the parent body (as implied by the texture and cooling history of EC 002). Such a body could escape global melting, and have melts generated at ≈ 1220 °C close to surface migrating over a few 100 kyr to the surface. This would allow the preservation of a primordial crust at the surface of the parent body of EC 002. Such andesitic crusts were perhaps quite common 4565 Myr ago. Possible parent bodies can be sought among available observations of asteroid surfaces. The spectral data obtained on EC 002 were compared with those obtained on asteroids, taking into account space weathering, the presence of olivine xenocrysts and residual mantle debris (SI Appendix). EC 002 is clearly distinguishable from all asteroid groups (42, 43), and no object with spectral characteristics similar to EC 002 has been identified to date. Remains of primordial andesitic crust are therefore not only rare in the meteorite record, they are also rare today in the asteroid belt. This suggests that the earliest differentiated protoplanets that populated the solar system, as well as most of their debris, were certainly destroyed or subsequently accreted to the growing rocky planets, making the discovery of meteorites originating from primordial crusts an exceptional occurrence.

**Materials and Methods**

Materials and methods are described in SI Appendix.

**Data Availability Statement.** All data discussed in the paper are given in SI Appendix, Tables S1–S7.

**Acknowledgments**

We thank X for the editorial handling, David Mittlefehldt and an anonymous reviewer for their constructive comments. We thank Addi Bischoff for showing us the probable occurrence of anorthoclase in EC 002. This work was supported by JSPS KAKENHI (Grant Number 19H01959), NIPR (Research Project KP307), ANR-15-CE31-0004-1 (ANR CRADLE), the UnivEarthS Labex program at Sorbonne Paris Cité (ANR-10-LABX-0023), the European Research Council under the H2020 framework program/ERC grant agreement no. 771691 (Solarys), the European Research Council (PHOTONIS project, grant agreement No. 695618), and by the "Laboratoire d'Excellence" LabexMER (ANR-10- LABX-19) and funded by grants from the French Government under the program "Investissements d'Avenir".

**References**

1. Meteoritical bulletin database, available at https://www.lpi.usra.edu/meteor/metbull.php
2. H. Y. McSween, et al., HED meteorites and their relationship to the geology of Vesta and the Dawn mission. *Space Sci. Reviews* 63, 141-174 (2011).
3. H. Downes, et al., Evidence from polymict ureilite meteorites for a disrupted and re-accreted single ureilite parent asteroid gardened by several distinct impactors. *Geochim. Cosmochim. Acta* 72, 4825-4844 (2008).
4. A. Yamaguchi, et al., A new source of basaltic meteorites inferred from Northwest Africa 011. *Science* 296, 334–336 (2002).
5. D.W. Mittlefehldt, Ibitira: a basaltic achondrite from a distinct parent asteroid and implications for the Dawn mission. *Meteorit. Planet. Sci.* 40, 665–677 (2005).
6. J.M.D. Day, et al., Early formation of evolved asteroidal crust. *Nature* 457(7226), 179-182 (2009).

7. C.K. Shearer, et al., Non-basaltic asteroidal melting during the earliest stages of solar system evolution. A view from Antarctic achondrites Graves Nunatak 06128 and 06129. *Geochim Cosmochim Acta* 74, 1172–1199 (2010).

8. A. Bischoff, et al., Trachyandesitic volcanism in the early Solar System. *Proc. Natl. Acad. Sci. U.S.A.* 111, 12689-12692 (2014).

9. P. Srinivasan, *et al.*, Silica-rich volcanism in the early solar system dated at 4.565 Ga. *Nature Commun.* 9, 3036 (2018).

10. T. Usui, J. H. Jones, D. W. Mittlefehldt, A partial melting study of an ordinary (H) chondrite composition with application to the unique achondrite Graves Nunataks 06128 and 06129. *Meteorit. Planet. Sci.* 50, 759–781 (2015).

11. N. G. Lunning et al., Partial melting of oxidized planetesimals: An experimental study to test the formation of oligoclase-rich achondrites Graves Nunataks 06128 and 06129. *Geochim. Cosmochim. Acta* 214, 73-85 (2017).

12. M. Collinet, T. L. Grove, Widespread production of silica- and alkali-rich melts at the onset of planetesimal melting. *Geochim. Cosmochim. Acta* 277, 334-357 (2020).

13. J.A. Barrat, et al., The formation of carbonates in the Tatahouine meteorite. *Science* 280, 412-414 (1998).

14. Meteoritical Bulletin, EC 002 entry, available at https://www.lpi.usra.edu/meteor/metbull.php?code=72475

15. D. Stöffler, C. Hamann, K. Metzler, Shock metamorphism of planetary silicate rocks and sediments: Proposal for an updated classification system. *Meteoritics Planet. Sci*. 53, 5-49 (2018).

16. J. Crank (1967) *The Mathematics of Diffusion.* Oxford University Press, London, 347 pp.

17. J. Ganguly, V. Tazzoli, $Fe^{2+}$-Mg interdiffusion in orthopyroxene: Retrieval from the data on intracrystalline exchange reaction, *Am. Mineral.* 79, 930-937 (1994).

18. R. W. Le Maitre et al., Igneous Rocks, a classification and glossary of terms. Cambridge University Press, ISBN 0 521 66215 X.

19. J.-A. Barrat, et al., Petrology and geochemistry of the unbrecciatted achondrite North West Africa 1240 (NWA 1240): an HED parent body impact melt. *Geochim. Cosmochim. Acta* 67, 3959-3970 (2003).

20. J.-A. Barrat, et al., Geochemistry of CI chondrites: Major and trace elements, and Cu and Zn isotopes. *Geochim Cosmochim Acta* 83, 79–92 (2012).

21. J.A. Barrat, et al., Evidence from Tm anomalies for non-CI refractory lithophile element proportions in terrestrial planets and achondrites. *Geochim. Cosmochim. Acta* 176, 1-17 (2016).

22. B. Jacobsen, et al., $^{26}$Al-$^{26}$Mg and $^{207}$Pb-$^{206}$Pb systematics of Allende CAIs: canonical solar initial $^{26}$Al/$^{27}$Al ratio reinstated. *Earth Planet. Sci. Lett.* 272, 353–364 (2008).

23. G.A. Brennecka, M. Wadhwa, Uranium isotope compositions of the basaltic angrite meteorites and the chronological implications for the early solar system. *Proc. Natl. Acad. Sci. USA* 109, 9299-9303 (2012).

24. M. Schiller, et al., Early accretion of protoplanets inferred from a reduced inner Solar System $^{26}$Al inventory. *Earth Planet. Sci. Lett.* 15, 45-54 (2015).

25. H. Busemann, O. Eugster, The trapped noble gas component in achondrites. *Meteoritics & Planetary Science* 37, 1865–1891 (2002).

26. T.S. Kruijer, et al. (2014) Protracted core formation and rapid accretion of protoplanets. *Science* 344, 1150–1154 (2014).

27. P. N. Peplowski, et al., Radioactive elements on Mercury's surface from MESSENGER: Implications for the planet's formation and evolution. *Science*, 333, 1850-1852 (2011).

28. F. Tera, et al., Comparative study of Li, Na, K, Rb, Cs, Ca, Sr and Ba abundances in achondrites and in Apollo 11 lunar samples. *Proc. Apollo 11 Lunar Sci. Conf.* 2, 1637–1657 (1970).

29. P.H. Warren, Stable-isotopic anomalies and the accretionary assemblage of the Earth and Mars: A subordinate role for carbonaceous chondrites. *Earth Planet. Sci. Lett.* 311, 93-100 (2011).

30. T. S. Kruijer, T. Kleine, Age and origin of IIE iron meteorites inferred from Hf-W chronology. *Geochim. Cosmochim Acta* 262, 92-103 (2019).

31. J. A. Barrat, et al., Partial melting of a C-rich asteroid : lithophile trace elements in ureilites. *Geochim. Cosmochim. Acta* 194, 163-178 (2016).

32. Y. Nakamuta, et al., Effect of NaCrSi$_2$O$_6$ component on Lindsley's pyroxene thermometer: an evaluation based on strongly metamorphosed LL chondrites. *Meteorit. Planet. Sci.* 52, 511-521 (2017).

33. A. Yamaguchi, et al., Experimental evidence of fast transport of trace elements in planetary crusts by high temperature metamorphism. *Earth Planet. Sci. Lett*. 368, 101-109 (2013).

34. H. Ono, Crustal evolution of asteroid Vesta as inferred from silica polymorphs in eucrites, University of Tokyo, 225 p (2020).

35. M. B. Olsen et al., Magnesium and $^{54}$Cr isotope compositions of carbonaceous chondrite chondrules - Insights into early disk processes. *Geochim. Cosmochim Acta* 191, 118-138 (2016).

36. M. Schiller, M. R. Handler, J. A. Baker. High-precision Mg isotopic systematics of bulk chondrites. *Earth Planet. Sci. Lett.* 297, 165–173 (2010).

37. T. H. Luu, et al., Bulk chondrite variability in mass independent magnesium isotope compositions - Implications for initial solar system $^{26}$Al/$^{27}$Al and the timing of terrestrial accretion. *Earth Planet. Sci. Lett.* 522, 166-175 (2019).

38. P. Koefoed, et al., U-Pb and Al–Mg systematics of the ungrouped achondrite Northwest Africa 7325. *Geochim. Cosmochim. Acta* 183, 31–45 (2016).

39. J. Wimpenny et al., Reassessing the origin and chronology of the unique achondrite Asuka 881394: Implications for distribution of $^{26}$Al in the early Solar System. Geochim. Cosmochim Acta 244, 478-501 (2019).

40. J. A. Barrat, et al., Crustal differentiation in the early solar system: clues from the unique achondrite Northwest Africa 7325 (NWA 7325). *Geochim. Cosmochim. Acta* 168, 280-292 (2015).

41. E. Kaminski et al., Early accretion of planetesimals unraveled by the thermal evolution of the parent bodies of magmatic iron meteorites. *Earth Planet. Sci. Lett.* 548, 116469 (2020).

42. F. E. DeMeo, R. P. Binzel, S. M. Slivan, S. J. Bus, An extension of the Bus asteroid taxonomy into the near-infrared. *Icarus* 202, 160–180 (2009).

43. F. E. DeMeo, B. Carry, The taxonomic distribution of asteroids from multi-filter all-sky photometric surveys. *Icarus* **226**, 723–741 (2013).

**Figures**

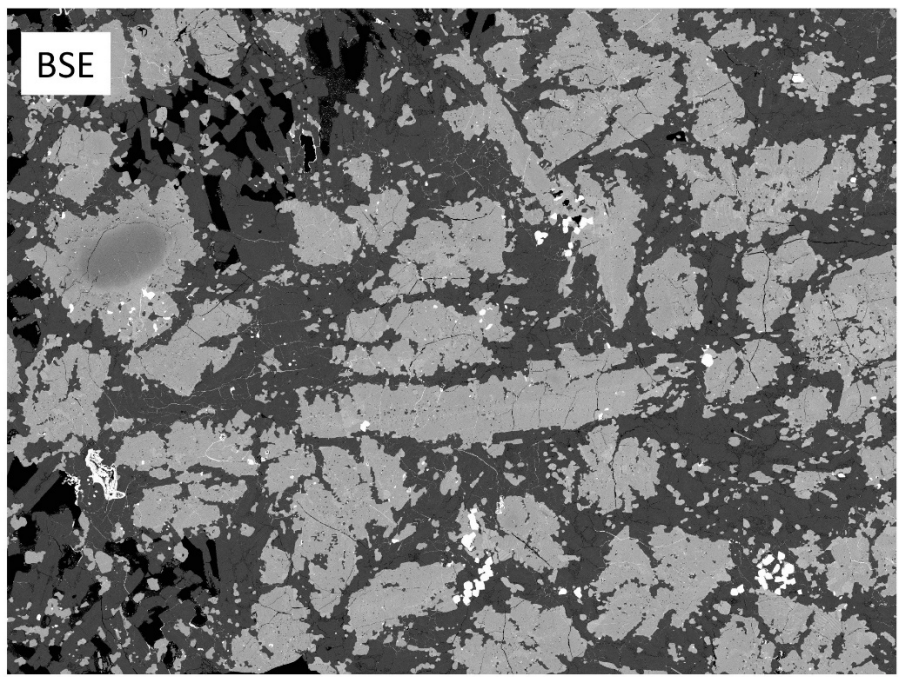
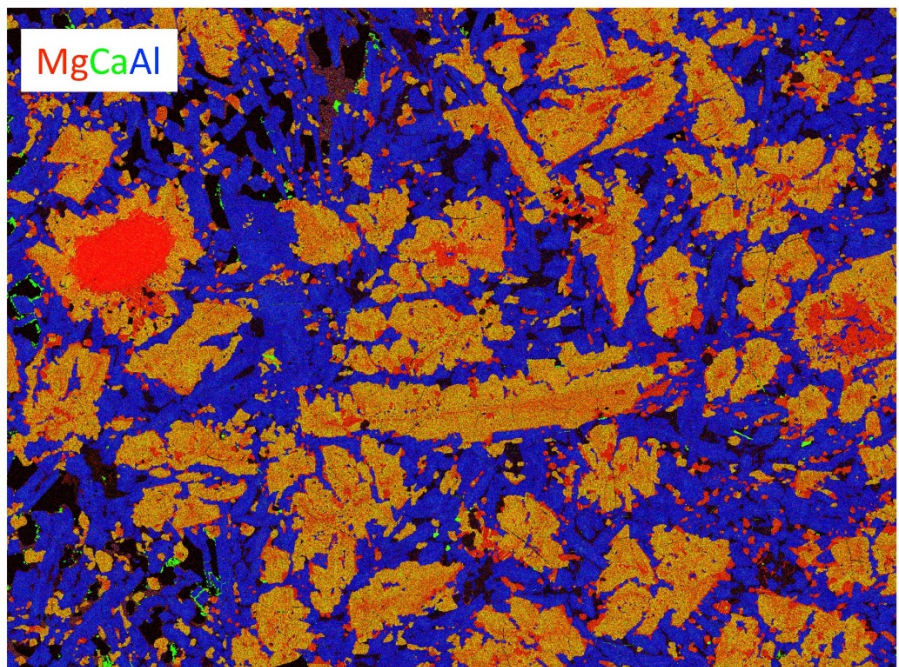

**Figure 1. (a)** Backscattered electron (BSE) image of a polished section of EC 002 (pyroxenes are gray to light gray, and feldspars are dark gray). **(b)** False-colored X-ray map of the same area. A Mg-rich orthopyroxene xenocryst (red) is seen in the left side of the image.

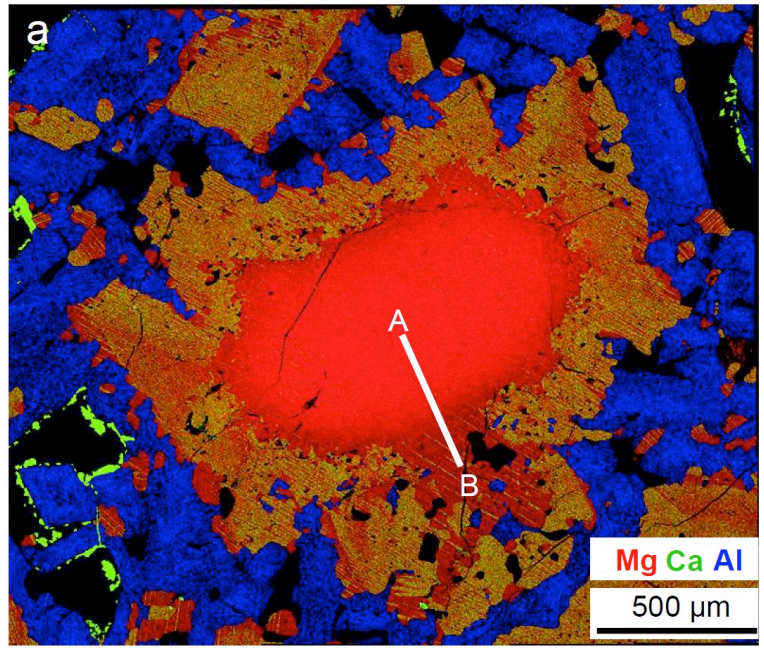

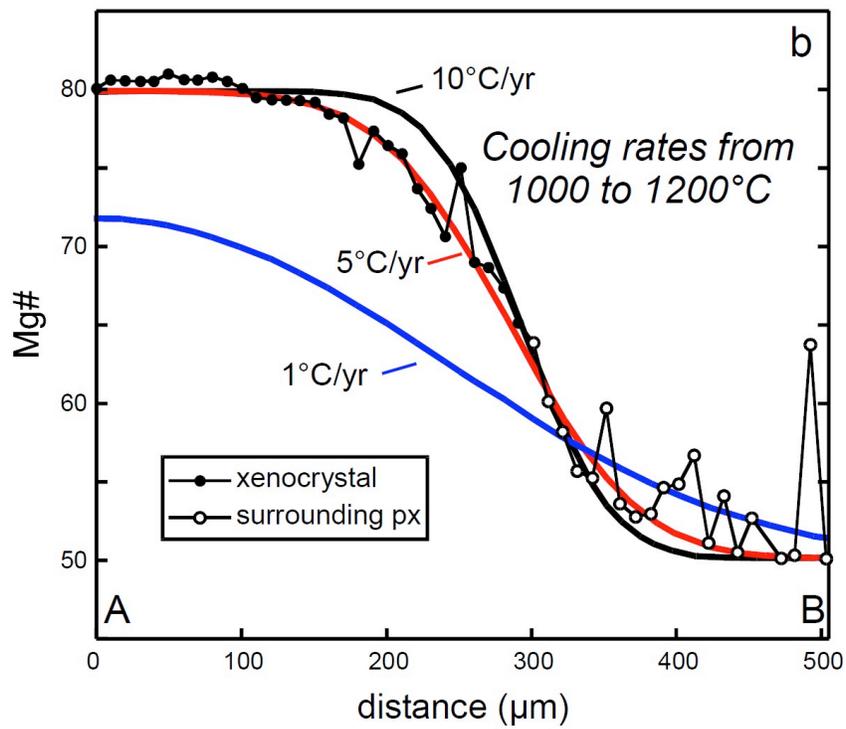

**Figure 2.** a/ false-colored X-ray map of a small orthopyroxene xenocryst (in red) mantled by groundmass pyroxenes; b/ Mg# profile across the boundary between xenocryst and groundmass pyroxene. We calculated

cooling rates assuming a compositionally homogeneous spherical xenocryst in a homogeneous medium (16). We used the interdiffusion coefficients of Fe-Mg in orthopyroxene (Mg# = 80) along the c and b directions (17). We chose the initial temperature of 1200°C, which is within the range of estimated crystallization temperatures and assumed constant cooling rates.

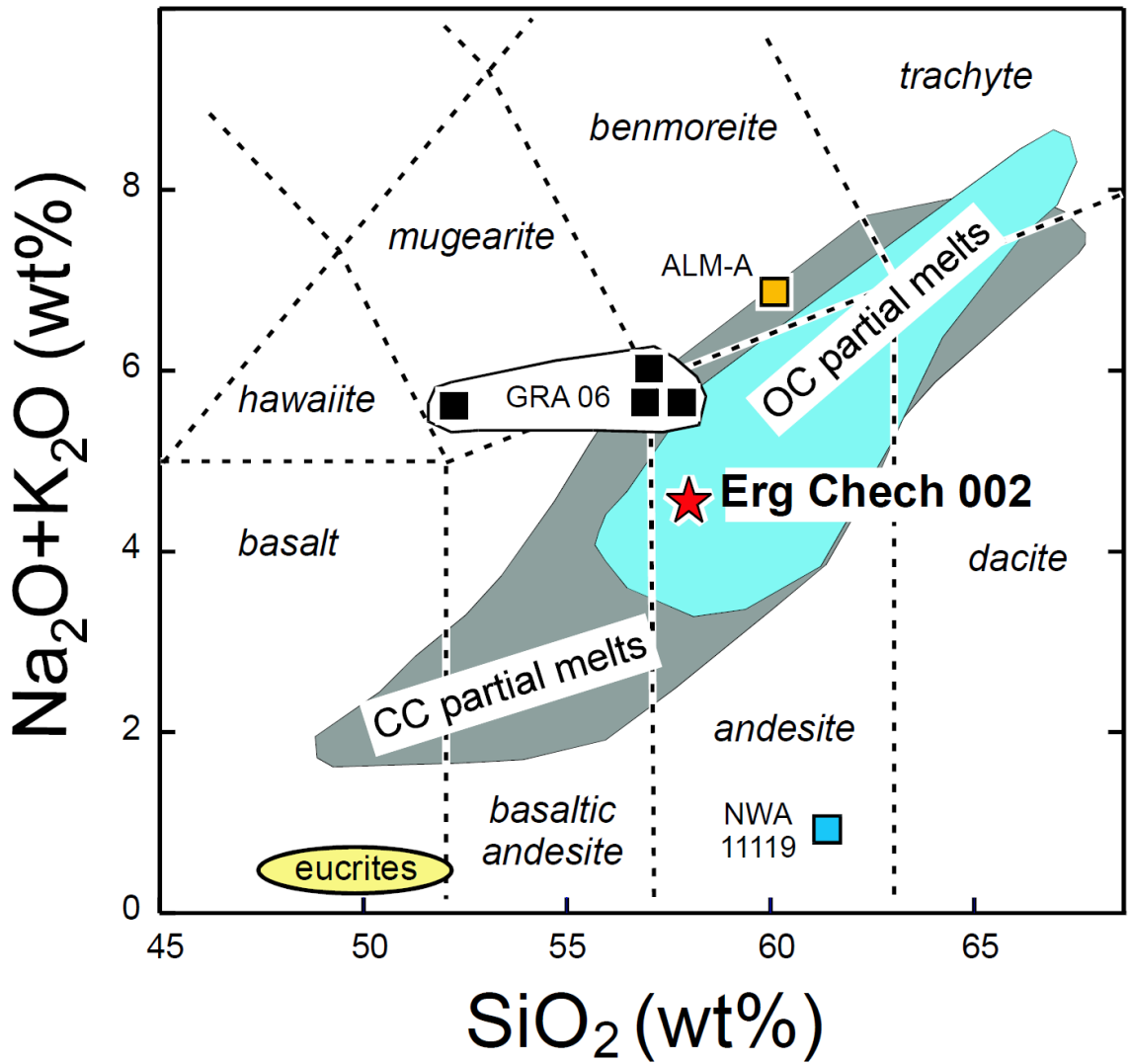

**Figure 3.** Plot of total alkalis vs. silica content showing the compositions EC 002, the other andesitic achondrites, eucrites, and the experimental melts obtained on chondritic systems by Collinet and Grove (12).

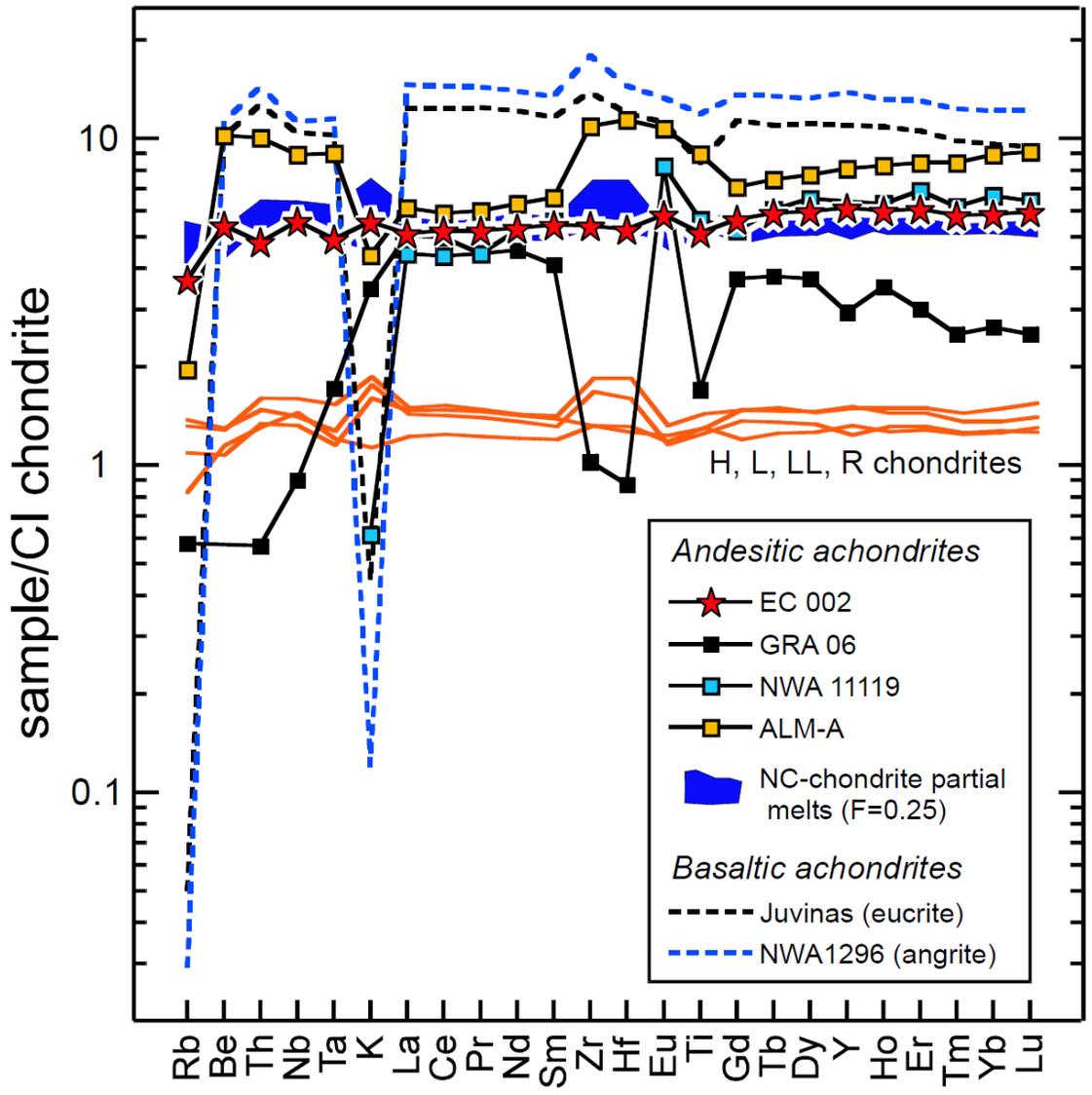

**Figure 4.** Comparison of CI–normalized trace element abundances in EC 002, other andesitic achondrites, representative basaltic achondrites and non-carbonaceous chondrites (6-9). The field of partial melts (F=25%) obtained from regular non-carbonaceous chondrites is shown for comparison. CI normalization values from (20).

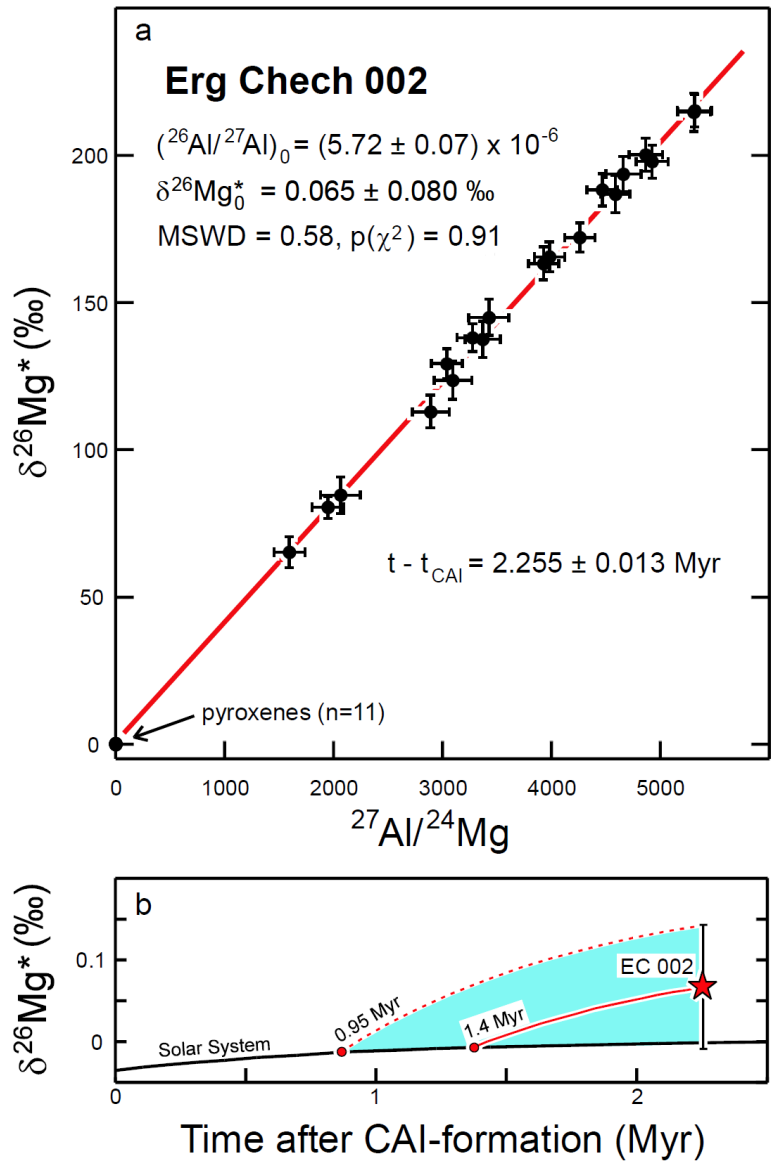

**Figure 5**. a/ The Al-Mg internal isochron defined by plagioclases and pyroxenes of EC 002. The error bars correspond to 2 x standard errors on the ratios. b/ Mg isotopic composition of EC 002 compared to the theoretical evolution of the Mg isotopic composition of the solar system (calculated for a chondritic $^{27}Al/^{24}Mg$ ratio of 0.101, a $(^{26}Al/^{27}Al)_0$ of 5.23 x $10^{-5}$, and a $\delta^{26}Mg*_0$ of -0.034‰ (22)).

# Supplementary Information

**Supplementary Information Text**

**A/ Petrography, mineralogy, geochemistry, geochronology**

**Materials and methods.**

*Petrography and phase compositions*

We studied two polished sections (~1.5 cm$^2$ each), one thick, one thin, made from the same slab. We examined these sections using an optical microscope equipped with both transmitted and reflected light, an electron microprobe analyzer (EPMA; JEOL JXA-8200) and a field emission scanning electron microscope (FE-SEM; JEOL JSM-7100F) equipped with an energy dispersive spectrometer (Oxford AZtec Energy) at the National Institute of Polar Research, Tokyo (NIPR). Pyroxene and spinel were analyzed using a current of 30 nA and a focused beam and plagioclase using 10 nA and a defocused beam (~3 µm in diameter) at 15 keV. For Mn analysis, we used a LIFH spectrometer. Count rates using the LIFH were increased by a factor of 3–4 compared to those of LIF. Bulk compositions of pyroxene are obtained from averaging 10-20 points across the grains using a beam diameter ~30 µm at 15 keV. Data were reduced using a ZAF correction program. Mineral phases were identified with a inVia Raman spectrometer at NIPR.

*Geochemistry*

Another slice weighing 1.09 g, was chosen after a thorough binocular examination of both sides because it contained no apparent pyroxene or olivine xenocrysts. It was crushed using a boron carbide mortar and pestle into a homogeneous fine grained powder in clean room conditions at Institut Universitaire Européen de la Mer (IUEM), Plouzané. Clear fragments (45 mg) of a large (> 1 cm) olivine crystal were leached in 6 N HCl (120°C, 30 minutes), rinsed 3 times, and dried. Major and trace elements were determined respectively by ICP-AES using a Horiba Jobin Yvon Ultima 2 spectrometer, and by ICP-SFMS using a Thermo scientific

ELEMENT XR spectrometer. We used the same procedures as Cotten et al. (1) for the major elements. The accuracy is better than 5 % for Na and P, and much better than 3 % for the other elements. For trace elements, we used the same procedure as Barrat et al. (2, 3). Based on standards and many sample duplicates, the precisions for abundances are in most cases much better than 5 % [two relative standard deviations (2 x RSD)]. The precisions for Eu/Eu* and Tm/Tm* ratios [X* is the expected X concentration for a smooth CI-normalised REE pattern, such that $Eu_n = (Sm_n \times Gd_n)^{1/2}$ and $Tm_n^* = (Er_n \times Yb_n)^{1/2}$] are respectively better than 3 % and 1.5 % (2 x RSD)].

## $^{26}Al$-$^{26}Mg$ dating

A thick slice (1 cm$^2$) was polished and mounted in a four-window ion probe holder. The Mg isotopic compositions and Al/Mg concentration ratios were measured with the CRPG-CNRS CAMECA ims 1280-HR2 ion microprobe.

*Ion probe settings for plagioclase*- Because of the low Mg contents (<0.03 wt% of MgO) of the plagioclases, the measurements were made in mono-collection mode using the central electron multiplier (EM) for $^{24}Mg^+$, $^{25}Mg^+$ and $^{26}Mg^+$ and the central Faraday cup (FC2) for $^{27}Al^+$. The samples were sputtered with a ~5 nA O$^-$ primary beam rastered at 15×15 μm. The transfer optic magnification was set at 100 μm to insure an efficient instrumental transmission and to fully fill the field aperture set at 2000 μm. The mass resolving power (MRP) was set at M/ΔM ~5000 in order to completely remove the $^{24}MgH^+$ interference on $^{25}Mg^+$. One measurement consisted of a 120 seconds pre-analysis sputtering to clean the sample surface and attain stable count rates on detectors followed by automatic secondary beam and energy centering, and 40 cycles with counting times during each cycle of 4, 5, 10, 10, 3 and 3 sec at masses 23.8 (background for EM), $^{24}Mg$, $^{25}Mg$, $^{26}Mg$, 26.8 (background FC2) and $^{27}Al$, respectively. The Mg isotopic compositions are given in delta' notation according to $\delta'^{25}Mg_X = \ln[(^{25}Mg/^{24}Mg)_X/(^{25}Mg/^{24}Mg)_{standard}] \times 1000$ (similarly for $^{26}Mg$) with $(^{25}Mg/^{24}Mg)_{standard} = 0.12663$ and $(^{26}Mg/^{24}Mg)_{standard} = 0.13932$. The $^{26}Mg$ excesses due to $^{26}Al$ decay are noted $\delta^{26}Mg^*$ with $\delta^{26}Mg^* = \delta'^{26}Mg - \delta'^{25}Mg/\beta$ (with β = 0.521 for an equilibrium mass fractionation of Mg isotopes as it is expected for the current object (4)). The choice of the mass fractionation law is not critical to the current data set since the degree of intrinsic mass fractionation is small and most likely due to unperfectly corrected matrix effects in plagioclases (Table S5) and because the Al/Mg ratios of plagioclase that control the isochron are high enough for the uncertainty introduced by the choice of fractionation law to be insignificant. Miyake-Jima plagioclase ($^{27}$Al/$^{24}$Mg=396.3) was used to calibrate the instrumental isotopic fractionation and the relative Al/Mg ion yield. Two sigma standard errors on the mean of ±0.40‰, ±0.31‰ and ±0.45‰ for δ'$^{25}$Mg, δ'$^{26}$Mg and δ$^{26}$Mg*, respectively, were obtained for the standard. The relative Al/Mg ion yield was determined to be 0.833±0.028. Typical counts rates for EC 002 plagioclases were 15000 cps, 2000 cps, 2500 cps and 4.2×10$^7$ cps $^{24}$Mg, $^{25}$Mg, $^{26}$Mg and $^{27}$Al, respectively. The errors reported for the measurements of the samples are two sigma errors calculated by summing in quadratic way the errors due to counting statistic in each point and the errors due to calibration of instrumental isotopic fractionation and Al/Mg ion yield (Table S5).

*Ion probe settings for pyroxene*- The measurements were made in multi-collection mode using four off-axis FCs (L1, C, H1 and H'2 for $^{24}$Mg+, $^{25}$Mg+, $^{26}$Mg+, $^{27}$Al+, respectively). The samples were sputtered with a ~6 nA O$^-$ primary beam rastered at 15×15 μm. The transfer optic magnification was set at 80 μm to insure an efficient instrumental transmission and to fully fill the field aperture set at 2500 μm. The mass resolving power (MRP) was set at M/ΔM ~5000 (slit 2) in order to completely remove the $^{24}$MgH$^+$ interference on $^{25}$Mg$^+$. One measurement consisted of a 90 seconds pre-analysis sputtering to clean the sample surface and attain stable count rates on detectors, during which offsets of FCs were measured, followed by automatic secondary beam and energy centering, and 40 cycles of 5 seconds integration time for data acquisition. Gold enstatite ($^{27}$Al/$^{24}$Mg=0.024) was used to calibrate the instrumental isotopic fractionation and the relative Al/Mg ion yield. Two sigma standard errors on the mean of ±0.04‰, ±0.05‰ and ±0.04‰ for δ'$^{25}$Mg, δ'$^{26}$Mg and δ$^{26}$Mg*, respectively, were obtained for the standard. The relative Al/Mg ion yield was determined to be 0.788±0.009. The errors reported for the measurements of the samples are two sigma errors calculated by summing in quadratic way the errors due to counting statistic in each point and the errors due to calibration of instrumental isotopic fractionation and Al/Mg ion yield (Table S6).

**Noble gas isotope analysis**

An 8.2 mg piece of the sample was loaded into a filament furnace consisting of 3 alumina-coated tungsten evaporation baskets. The sample was placed directly into one of the baskets whilst two empty baskets served to calculate the blank contribution. The full analytical procedure has been previously documented in Broadley et al. (5). In brief, the furnace was pumped while baking at 150°C for 48 hours to remove adsorbed atmospheric gases from the walls of the furnace and the sample. The furnace was then pumped for a further 48 hours to ensure low blank levels. Gases were extracted from the sample over five temperature steps at 600°C, 800°C, 1000°C, 1200°C, and 1400°C. The majority of the He (90%) was released at the 800°C extraction step. For both Ne (84%) and Ar (91%), the majority was released at 1200°C. Negligible amounts of gas were released at 1400°C, and inspection of the sample after analysis confirmed it had been completely vaporized.

The extracted He, Ne and Ar were purified, cryo-separated and analyzed using a Helix-MC mass spectrometer (5). Blanks were analyzed following exactly the same protocol as samples, and for the major extraction step, the $^4$He, $^{21}$Ne and $^{40}$Ar blanks represent 0.01%, 0.01% and 0.14% of the sample respectively. Blank contributions for $^{36}$Ar were significantly higher given the very low concentration of $^{36}$Ar in the sample, with blank contribution reaching a maximum of 50% for the 1000°C temperature step. Noble gas abundances and isotope ratios are reported in Table S7

Exposure ages were calculated for $^3$He and $^{21}$Ne following Leya et al., (6), using target element abundances from the bulk chemical composition shown in Table S4 and a shielding depth correction factor of 20 cm. The

K-Ar age was calculated following Kelley (7), assuming that 100% of the measured $^{40}$Ar is radiogenic. A correction for atmospheric $^{40}$Ar results in a change of <1% and is insignificant compared to the uncertainty already associated with the measurement. The trapped $^{36}$Ar component was calculated using simple 2-endmember mixing, using a theoretical cosmogenic $^{38}$Ar/$^{36}$Ar of 1.438, and an assumed trapped $^{38}$Ar/$^{36}$Ar of 0.188 (atmospheric, but also similar to other trapped planetary phases (8)). This gives the fraction of trapped $^{36}$Ar as 21%, and a total trapped abundance of 4.3 x10$^{-13}$ mol/g.

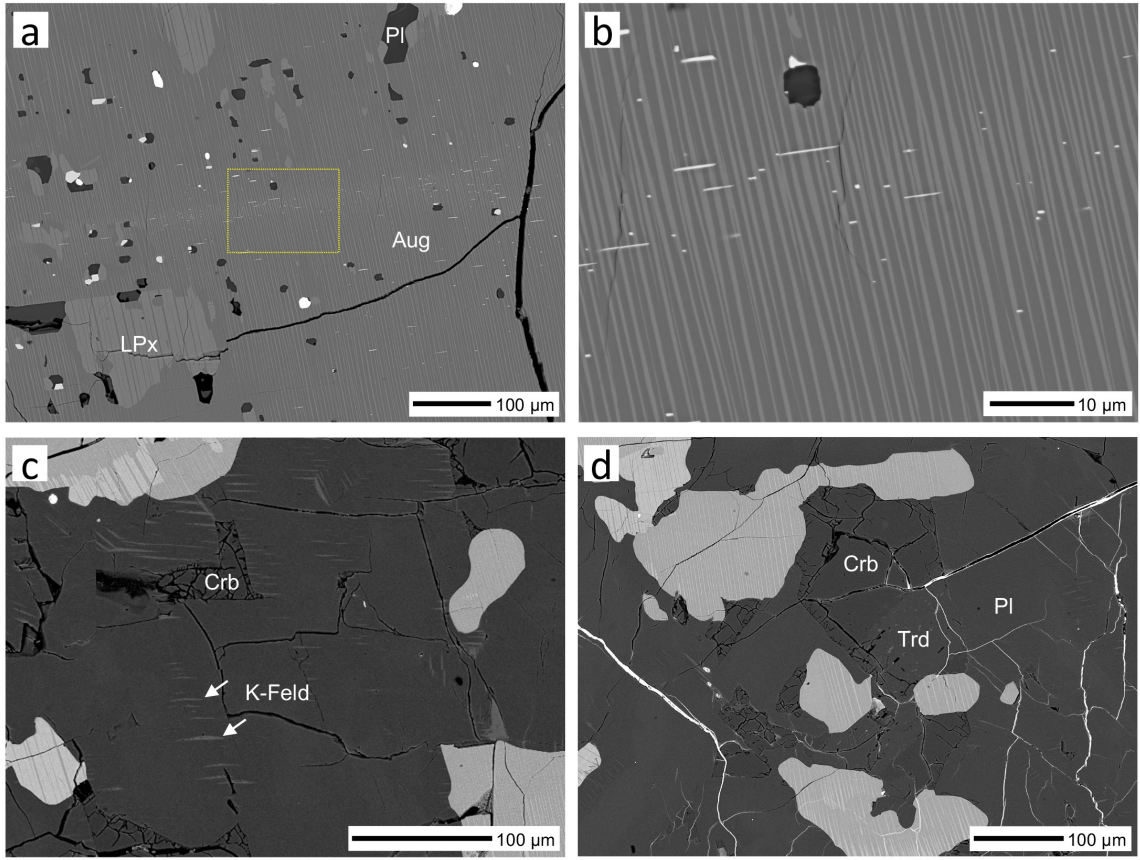

**Figure S1.** Back-scattered electron (BSE) images of EC 002. (a) Groundmass pyroxene composed of relict augite (Aug) and minor low-Ca pyroxene (LPx). Small inclusions of plagioclase (dark gray) and oxide minerals (white) occur. (b) Enlarged view of the dotted square in (a). Notice closely-spaced augite (medium gray) and low Ca-pyroxene lamellae (light gray). (c) Thin lamellae of K-feldspar (light gray) in plagioclase. (d) Cristobalite (Crb) and tridymite (Trd). Cristobalite is slightly fractured. Tridymite is slightly brighter under BSE and has fine inclusions of plagioclase.

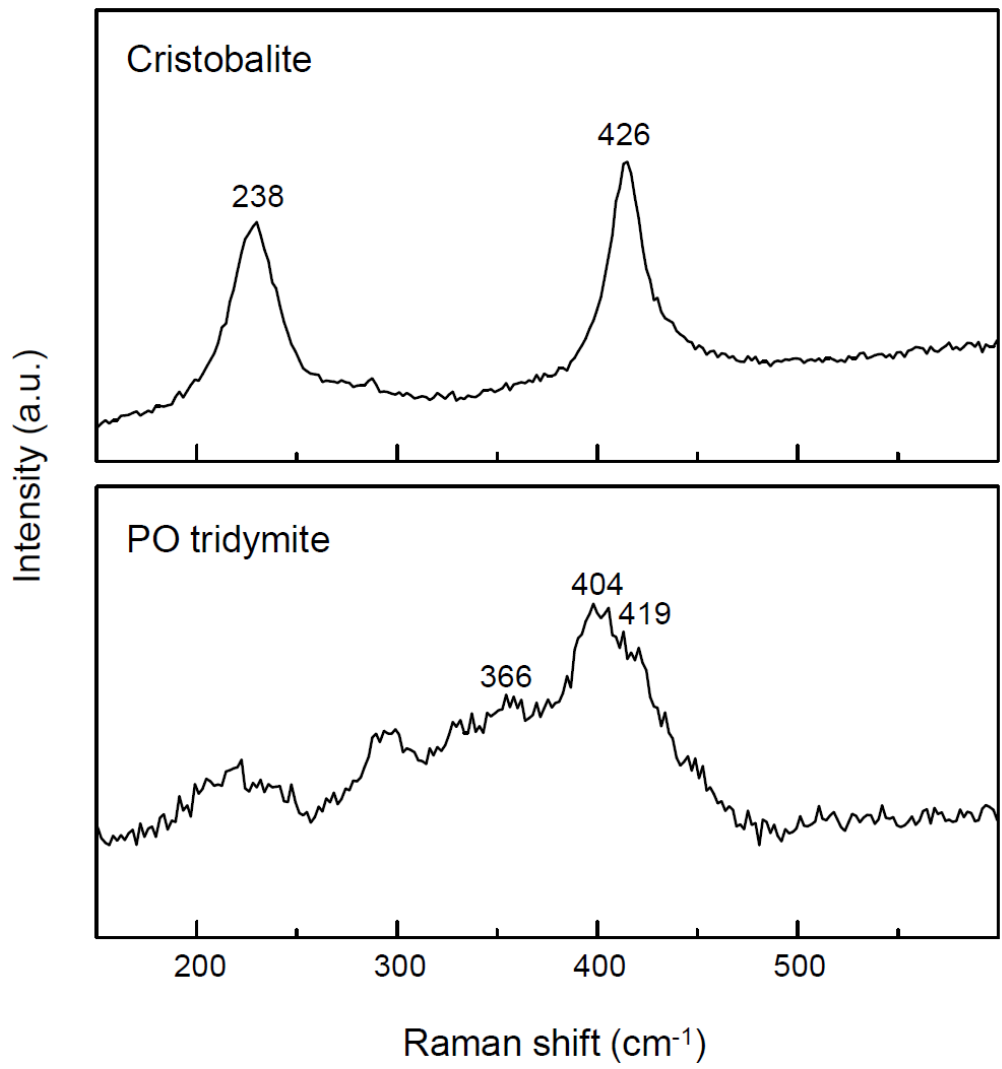

**Figure S2.** Representative Raman spectra of silica in EC 002.

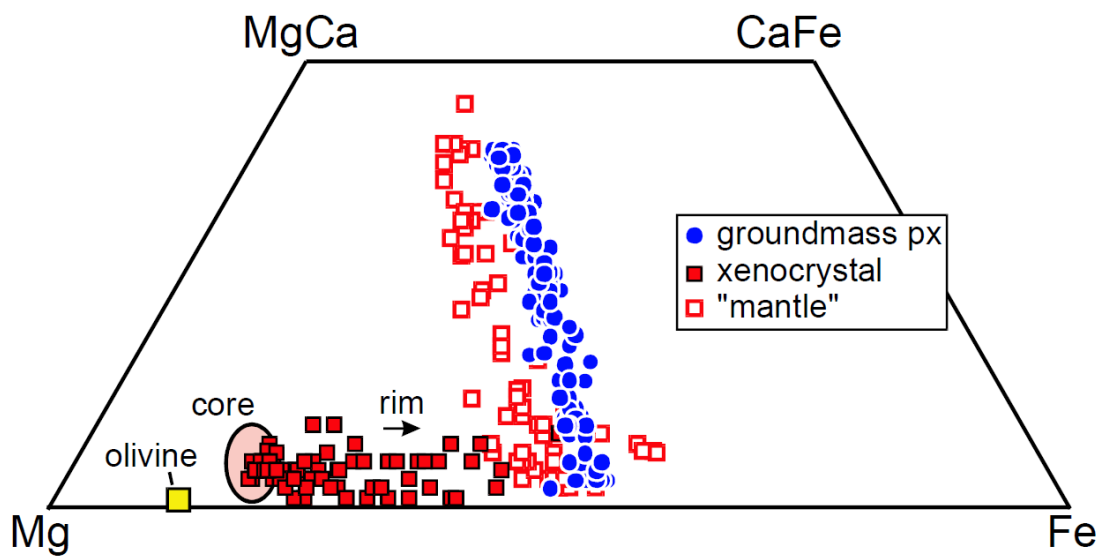

**Figure S3.** Olivine megacryst and pyroxene compositions. Groundmass pyroxenes are low-Ca pyroxene and augite. The intermediate Wo contents are due to incomplete spatial resolutions of electron microprobe.

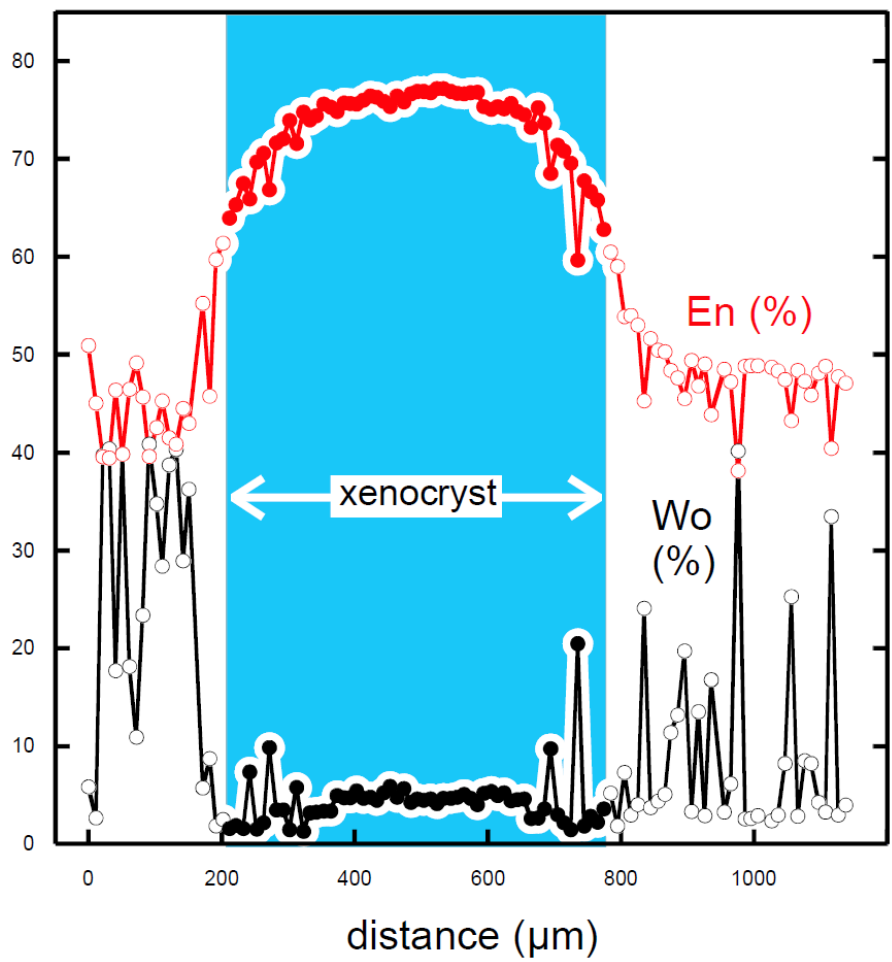

**Figure S4**. Chemical zoning (enstatite and wollastonite component, in mol%) of a small pyroxene xenocryst mantled by groundmass pyroxenes.

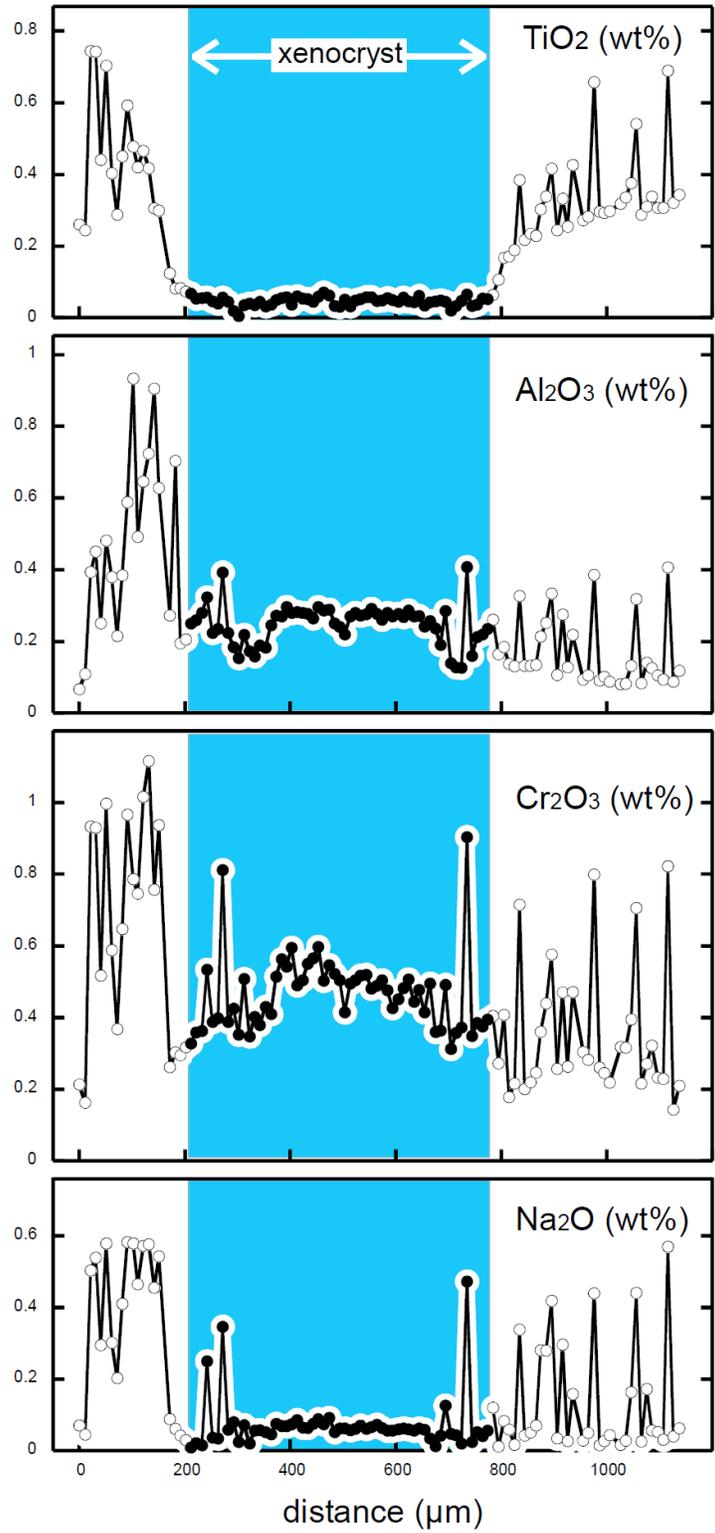

**Figure S5**. Chemical zoning (TiO$_2$, Al$_2$O$_3$, Cr$_2$O$_3$ and Na$_2$O in wt%) of a small pyroxene xenocryst mantled by groundmass pyroxenes.

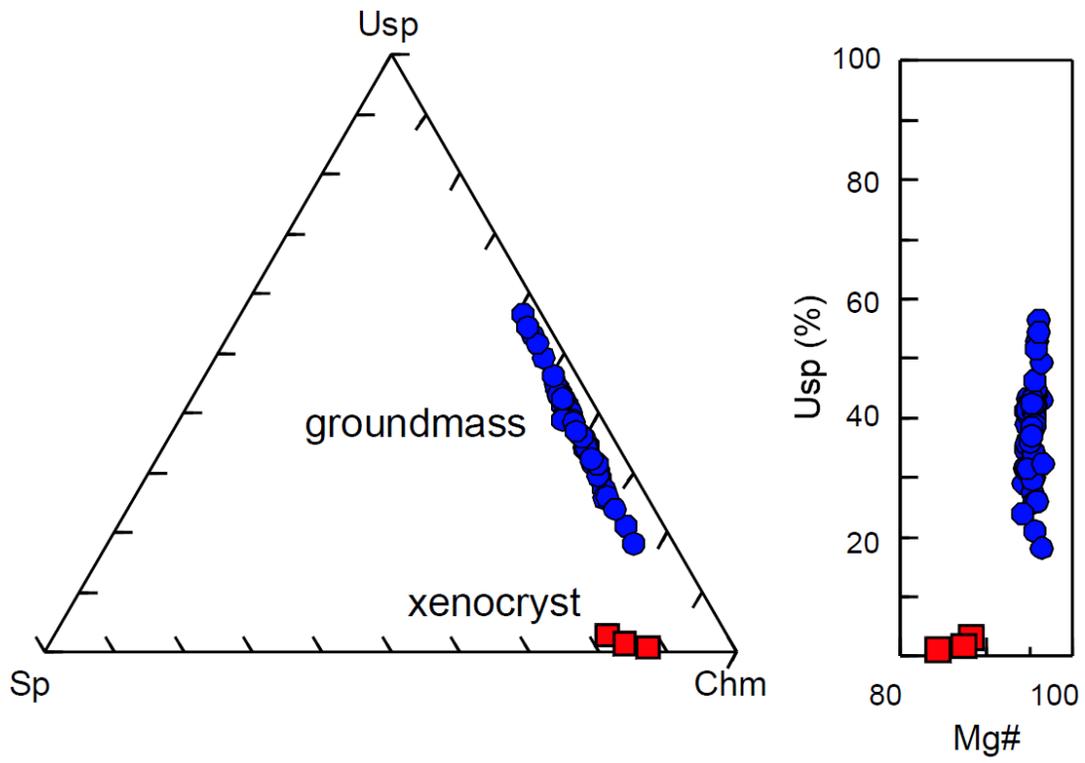

**Figure S6**. Spinel compositions.

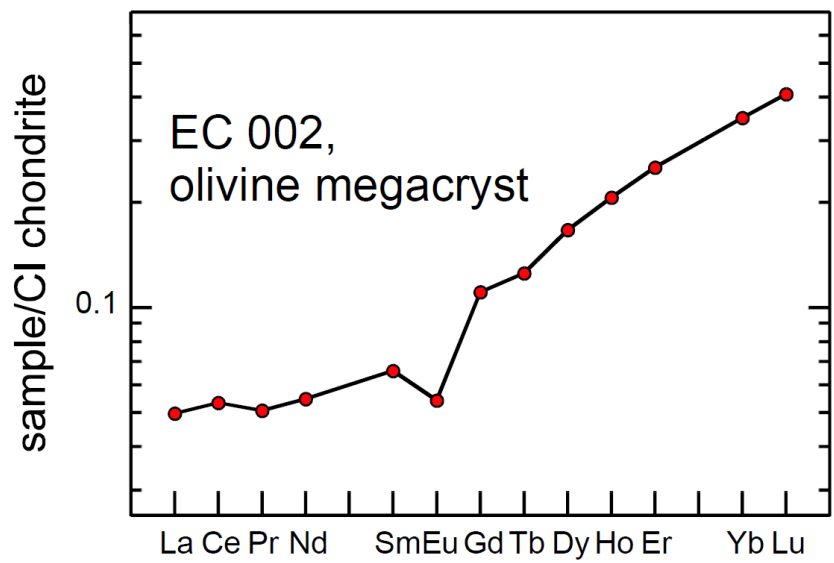

**Figure S7.** REE pattern of an olivine megacryst found in EC 002.

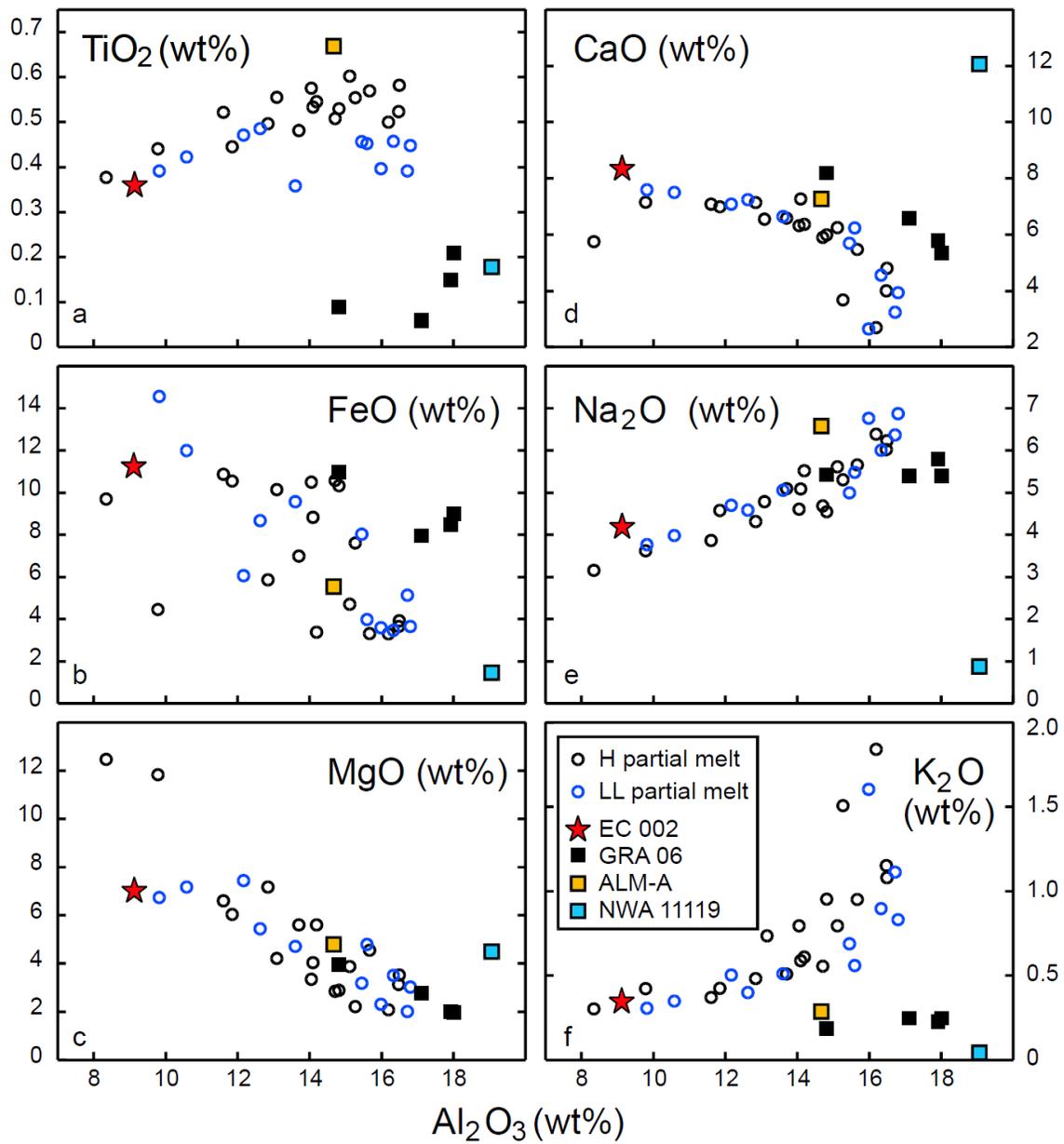

**Figure S8**. Compositions of EC 002 (this work) and other andesitic achondrites (9-11) compared with experimental partial melts from system with H or LL-chondrite compositions (12).

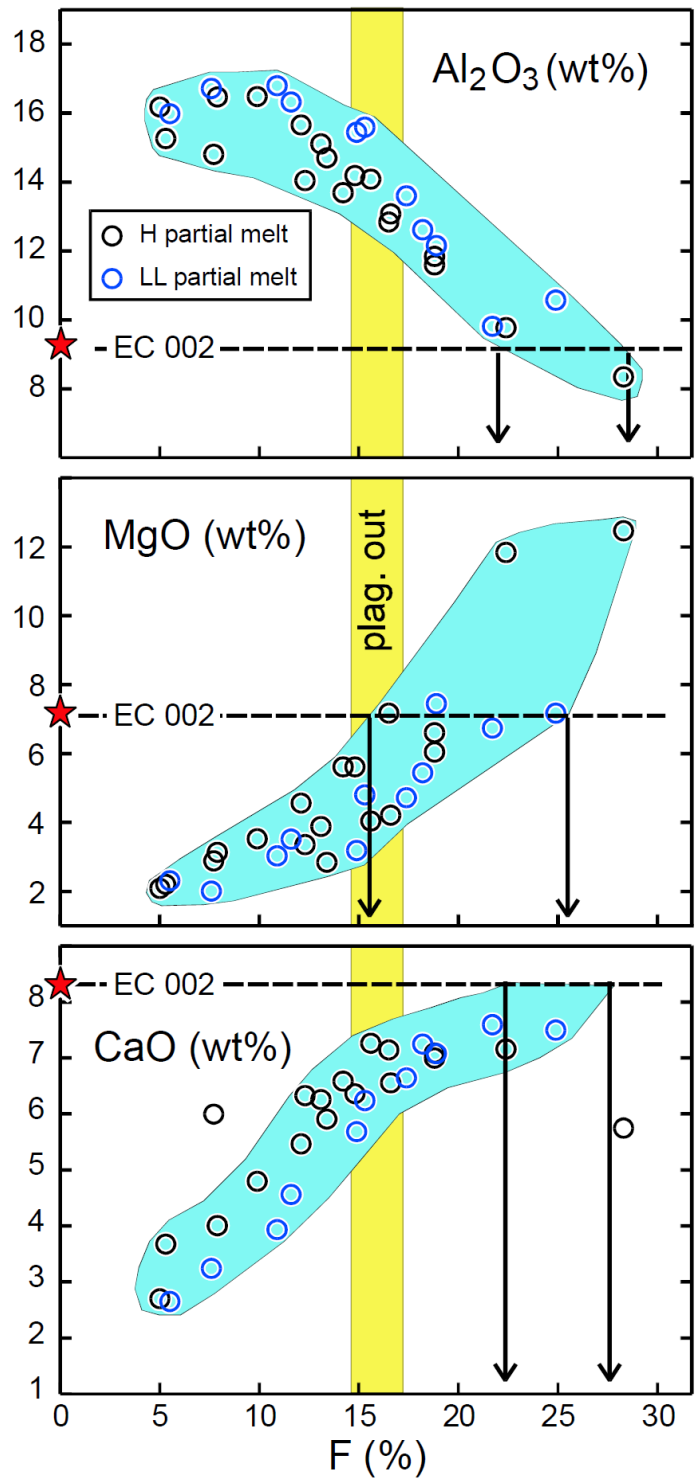

**Figure S9**. Compositions of EC 002 (this work) compared with experimental partial melts from system with H or LL-chondrite compositions (12). The composition of EC 002 suggests high melting degrees (F) around 25%.

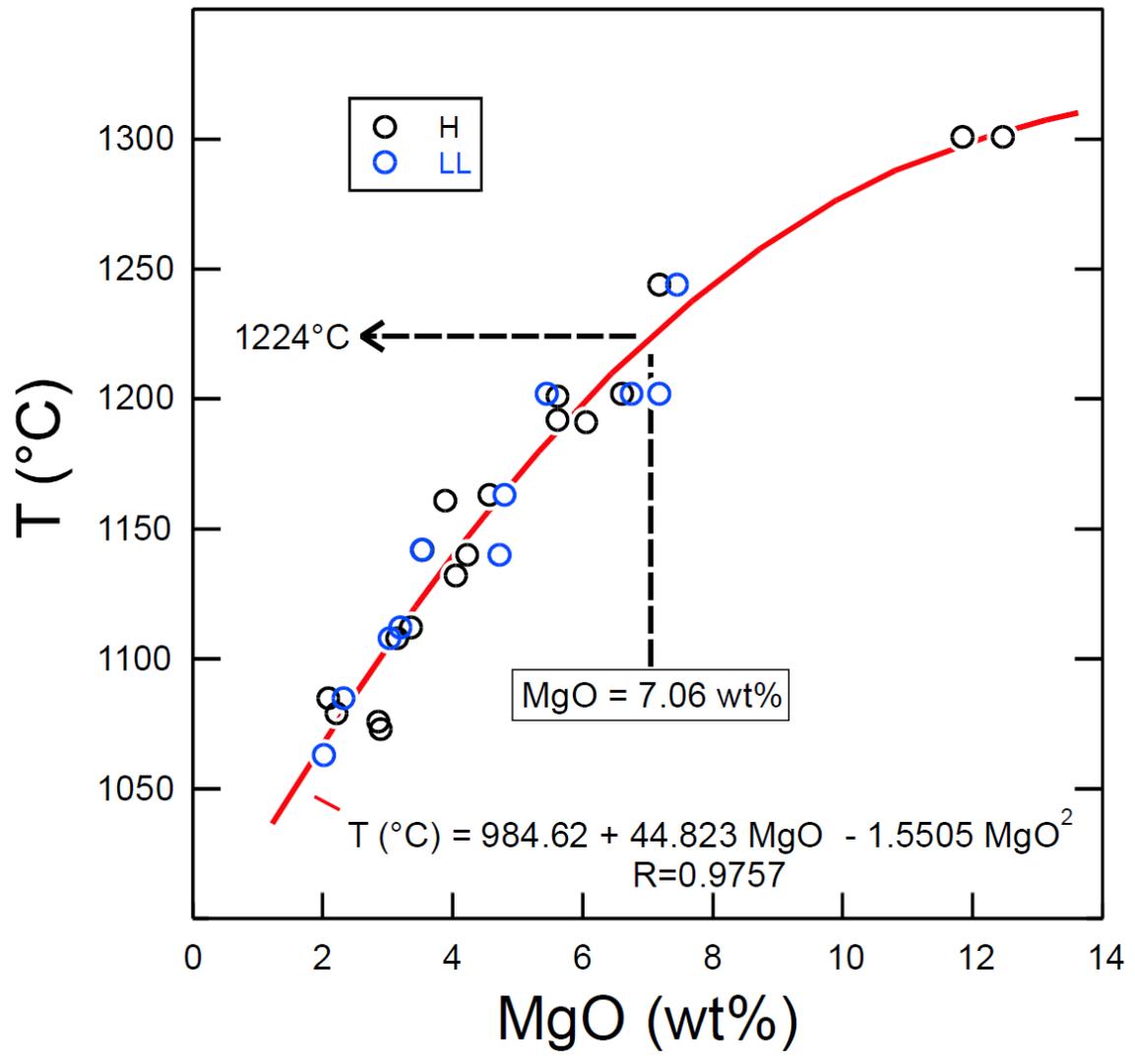

**Figure S10**. Liquidus temperature vs. MgO (wt%) for experimental melts obtained by partial melting from system with H or LL-chondrite compositions (12). The data are well described by a simple polynomial expression with errors < 20°C. A liquidus temperature of 1220°C is estimated for EC 002 (MgO=7.06 wt%).

Table S1. Representative compositions of pyroxenes and oxides in EC 002 (in wt%, n.d.: not determined).

| | $SiO_2$ | $TiO_2$ | $Al_2O_3$ | $Cr_2O_3$ | FeO | MnO | MgO | CaO | $Na_2O$ | $V_2O_3$ | NiO | ZnO | Total | | | |
|---|---|---|---|---|---|---|---|---|---|---|---|---|---|---|---|---|
| **Pyroxene** | | | | | | | | | | | | | | | | |
| Groundmass | | | | | | | | | | | | | | Wo | En | Fs |
| Low-Ca | 51.24 | 0.32 | 0.09 | 0.12 | 30.4 | 1.34 | 15.2 | 1.44 | 0.03 | n.d. | n.d. | n.d. | 100.16 | 3.12 | 45.57 | 51.31 |
| High-Ca | 51.87 | 0.27 | 0.96 | 1.17 | 13.6 | 0.64 | 12.1 | 18.6 | 0.54 | n.d. | n.d. | n.d. | 99.66 | 40.46 | 36.50 | 23.03 |
| Xenocryst | | | | | | | | | | | | | | | | |
| Core | 55.43 | 0.06 | 0.28 | 0.52 | 12.2 | 0.59 | 28.5 | 2.40 | 0.06 | n.d. | n.d. | n.d. | 100.05 | 4.65 | 76.89 | 18.47 |
| Rim | 51.91 | 0.24 | 0.13 | 0.22 | 27.2 | 1.25 | 17.1 | 2.06 | 0.05 | n.d. | n.d. | n.d. | 100.16 | 4.38 | 50.46 | 45.16 |
| **Spinel** | | | | | | | | | | | | | | Usp | Sp | Cm |
| Low-Ti | 0.04 | 6.38 | 2.59 | 50.74 | 36.6 | 1.25 | 0.77 | 0.19 | <0.01 | 0.44 | <0.03 | <0.02 | 98.61 | 18.19 | 5.79 | 76.02 |
| High-Ti | 0.04 | 20.1 | 1.19 | 27.67 | 48.8 | 0.91 | 1.13 | <0.01 | <0.01 | 0.23 | <0.03 | 0.04 | 99.84 | 56.47 | 2.63 | 40.90 |
| **Ilmenite** | | | | | | | | | | | | | | | | |
| Average | <0.02 | 53.9 | <0.01 | 0.20 | 44.13 | 1.07 | 1.64 | <0.01 | <0.01 | <0.03 | <0.03 | 0.01 | 100.99 | | | |
| *1SD, n=9* | | *0.42* | | *0.08* | *0.58* | *0.07* | *0.20* | | | | | *0.01* | | | | |

Table S2. Compositions of bulk pyroxenes (in wt%, n= number of analyses).

| | n | SiO$_2$ | TiO$_2$ | Al$_2$O$_3$ | Cr$_2$O$_3$ | FeO | MnO | MgO | CaO | Na$_2$O | Total | Wo | En | Fs |
|---|---|---|---|---|---|---|---|---|---|---|---|---|---|---|
| **Groundmass pyrox.** | | | | | | | | | | | | | | |
| P3-1 | 21 | 52.0 | 0.40 | 0.68 | 0.83 | 17.6 | 0.79 | 12.1 | 15.1 | 0.47 | 100.0 | 32.99 | 36.93 | 30.08 |
| | *1SD* | *0.3* | *0.17* | *0.27* | *0.15* | *1.2* | *0.05* | *0.4* | *1.0* | *0.10* | | | | |
| P3-2 | 44 | 51.6 | 0.59 | 0.51 | 0.73 | 19.2 | 0.85 | 12.3 | 13.7 | 0.40 | 99.9 | 29.99 | 37.33 | 32.69 |
| | *1SD* | *0.6* | *0.09* | *0.15* | *0.10* | *1.7* | *0.07* | *0.4* | *1.7* | *0.05* | | | | |
| P-4 | 15 | 51.9 | 0.34 | 0.89 | 0.86 | 17.9 | 0.79 | 12.2 | 14.2 | 0.51 | 99.7 | 31.47 | 37.53 | 31.00 |
| | *1SD* | *1.1* | *0.09* | *0.50* | *0.17* | *2.9* | *0.11* | *0.7* | *2.9* | *0.24* | | | | |
| P-5 | 20 | 51.9 | 0.35 | 1.88 | 1.16 | 18.5 | 0.81 | 11.9 | 12.0 | 0.92 | 99.4 | 27.71 | 38.52 | 33.77 |
| | *1SD* | *2.7* | *0.28* | *2.92* | *1.48* | *3.1* | *0.14* | *1.8* | *3.4* | *1.31* | | | | |
| P-6 | 20 | 51.9 | 0.42 | 0.76 | 0.90 | 18.0 | 0.82 | 12.3 | 14.3 | 0.47 | 99.9 | 31.51 | 37.61 | 30.88 |
| | *1SD* | *0.6* | *0.13* | *0.29* | *0.30* | *2.0* | *0.08* | *0.5* | *2.0* | *0.13* | | | | |
| P-7 | 20 | 51.8 | 0.38 | 0.85 | 0.82 | 17.7 | 0.79 | 12.2 | 14.4 | 0.48 | 99.5 | 31.51 | 37.61 | 30.88 |
| | *1SD* | *0.9* | *0.10* | *0.48* | *0.13* | *1.4* | *0.06* | *0.5* | *1.1* | *0.18* | | | | |
| **Xenocryst core** | | | | | | | | | | | | | | |
| | 6 | 55.54 | 0.03 | 0.28 | 1.02 | 12.15 | 0.58 | 28.21 | 2.40 | 0.06 | 100.2 | 4.69 | 76.77 | 18.54 |
| | *1SD* | *0.48* | *0.01* | *0.03* | *0.45* | *0.25* | *0.01* | *0.44* | *0.52* | *0.02* | | | | |

**Table S3**. Chemical compositions of feldspars and silica (in wt%).

| | $SiO_2$ | $TiO_2$ | $Al_2O_3$ | $Cr_2O_3$ | FeO | MnO | MgO | CaO | $Na_2O$ | $K_2O$ | Total | Or | Ab | An |
|---|---|---|---|---|---|---|---|---|---|---|---|---|---|---|
| **Feldspar** | | | | | | | | | | | | | | |
| Low-Na | 63.2 | <0.03 | 23.0 | <0.03 | 0.08 | <0.03 | <0.03 | 4.45 | 8.63 | 0.47 | 99.8 | 2.69 | 75.75 | 21.55 |
| High-Na | 66.7 | 0.09 | 20.3 | <0.03 | 0.17 | <0.03 | <0.03 | 1.40 | 9.90 | 1.23 | 99.8 | 7.04 | 86.22 | 6.74 |
| High-K | 64.3 | 0.12 | 18.8 | 0.03 | 0.31 | <0.03 | <0.03 | 0.87 | 1.26 | 14.3 | 100.0 | 84.39 | 11.31 | 4.30 |

**Silica minerals**

| | | | | | | | | | | |
|---|---|---|---|---|---|---|---|---|---|---|
| 1 | 95.9 | 0.21 | 2.48 | 0.02 | 0.04 | <0.03 | <0.03 | 0.09 | 1.20 | 0.01 | 100.0 |
| 2 | 95.8 | 0.20 | 1.87 | <0.03 | 0.05 | <0.03 | <0.03 | 0.04 | 0.77 | 0.42 | 99.2 |

Table S4. Bulk chemical composition of EC 002 and of an olivine megacryst (oxydes in wt%, trace elements in µg/g, n.d. = not determined).

|  | W.R. | olivine |  | W.R. | olivine |  | W.R. | olivine |
|---|---|---|---|---|---|---|---|---|
| $SiO_2$ | 58.01 | (38.53) | Li | 3.59 | n.d. | Sm | 0.833 | 0.0101 |
| $TiO_2$ | 0.36 | 0.09 | Be | 0.122 | 0.007 | Eu | 0.342 | 0.00318 |
| $Al_2O_3$ | 9.12 | 1.66 | P | 216 | < 10 | Gd | 1.16 | 0.0228 |
| $Cr_2O_3$ | 0.42 | 1.00 | K | 3057 | 102 | Tb | 0.222 | 0.00470 |
| FeO | 11.22 | 11.11 | Sc | 30.49 | 9.90 | Dy | 1.52 | 0.0423 |
| MnO | 0.47 | 0.46 | Ti | 2309 | 303 | Ho | 0.339 | 0.0117 |
| MgO | 7.06 | 45.88 | V | 97.0 | 98.7 | Er | 1.01 | 0.0418 |
| CaO | 8.31 | 1.27 | Mn | 3289 | 3703 | Tm | 0.152 | n.d. |
| $Na_2O$ | 4.20 | n.d. | Co | 5.85 | 10.15 | Yb | 0.984 | 0.0586 |
| $K_2O$ | 0.34 | n.d. | Ni | 18.49 | 1.31 | Lu | 0.147 | 0.0100 |
| $P_2O_5$ | 0.06 | n.d. | Cu | 1.35 | 0.05 | Hf | 0.563 | 0.00491 |
| total | 99.58 | (100) | Zn | 0.44 | 0.09 | Ta | 0.0726 | n.d. |
|  |  |  | Ga | 2.60 | 0.13 | W | 0.022 | n.d. |
| Mg# | 52.87 | 88.04 | Rb | 8.60 | 4.88 | Pb | 0.090 | 0.0048 |
|  |  |  | Sr | 58.45 | 0.44 | Th | 0.136 | n.d. |

| CIPW norm | | Y | 9.55 | 0.357 | U | 0.109 | 0.001 |
|---|---|---|---|---|---|---|---|
| qz | 2.51 | Zr | 18.96 | 0.301 | | | |
| pl | 40.57 | Nb | 1.61 | 0.040 | $La_n/Sm_n$ | 0.94 | 0.75 |
| or | 2.01 | Cs | 0.293 | 0.284 | $Gd_n/Lu_n$ | 0.95 | 0.27 |
| di | 29.78 | Ba | 47.51 | 0.198 | Eu/Eu* | 1.05 | 0.63 |
| hy | 23.27 | La | 1.20 | 0.0117 | Tm/Tm* | 0.973 | - |
| il | 0.68 | Ce | 3.14 | 0.0320 | Zr/Hf | 33.68 | 61.30 |
| ap | 0.14 | Pr | 0.475 | 0.00461 | Nb/Ta | 22.18 | - |
| chr | 0.62 | Nd | 2.46 | 0.0254 | | | |

**Table S5.** Al-Mg data for EC 002 plagioclases.

| | $^{27}Al/^{24}Mg$ | 2SE | $\delta^{25}Mg$ | 2SE | $\delta^{26}Mg$ | 2SE | $\delta^{26}Mg*$ | 2SE |
|---|---|---|---|---|---|---|---|---|
| plag#1 | 4262 | 136 | 16 | 2.3 | 202.7 | 2.2 | 172 | 5.0 |
| plag#2 | 1597 | 142 | 7.2 | 2.3 | 79.0 | 2.8 | 65.2 | 5.2 |
| plag#3 | 4867 | 152 | 12.3 | 2.4 | 223.8 | 3.2 | 200.2 | 5.7 |
| plag#5 | 3427 | 182 | 10.1 | 2.4 | 164.3 | 3.7 | 144.9 | 6.0 |
| plag#6 | 2066 | 182 | 9.7 | 2.4 | 103.1 | 4.1 | 84.6 | 6.2 |
| plag#7 | 3983 | 138 | 11.9 | 2.3 | 188.4 | 2.2 | 165.5 | 5.1 |
| plag#8 | 4661 | 161 | 11.2 | 2.5 | 215.1 | 3.2 | 193.6 | 5.9 |
| plag#9 | 3372 | 162 | 11.9 | 2.5 | 160.4 | 3.9 | 137.5 | 6.2 |
| plag#10 | 2893 | 169 | 11.0 | 2.5 | 134.1 | 2.7 | 112.9 | 5.5 |
| plag#11 | 3039 | 141 | 7.0 | 2.4 | 142.7 | 2.1 | 129.2 | 5.1 |
| plag#14 | 4586 | 136 | 10.9 | 3.0 | 207.8 | 2.5 | 186.9 | 6.3 |
| plag#15 | 5316 | 158 | 11.5 | 2.5 | 237.3 | 2.4 | 215.2 | 5.5 |
| plag#16 | 4926 | 147 | 13.1 | 2.4 | 223.1 | 3.3 | 197.9 | 5.6 |
| plag#17 | 3275 | 141 | 11.2 | 2.2 | 159.6 | 2 | 138.1 | 4.7 |
| plag#18 | 4469 | 141 | 10.0 | 2.6 | 207.4 | 1.9 | 188.3 | 5.5 |
| plag#19 | 5310 | 152 | 13.6 | 3.0 | 240.8 | 2.9 | 214.6 | 6.5 |
| plag#20 | 3930 | 140 | 10.5 | 2.6 | 183.4 | 2.1 | 163.3 | 5.5 |
| plag#21 | 3097 | 172 | 10.4 | 2.7 | 143.5 | 3.8 | 123.6 | 6.5 |
| plag#22 | 1949 | 145 | 5.8 | 1.5 | 91.7 | 2.4 | 80.5 | 3.7 |

Table S6. Al-Mg data for EC 002 pyroxenes.

| | $^{27}Al/^{24}Mg$ | 2SE | $\delta^{25}Mg$ | 2SE | $\delta^{26}Mg$ | 2SE | $\delta^{26}Mg^*$ | 2SE |
|---|---|---|---|---|---|---|---|---|
| px #1 | 0.0362 | 0.0001 | 1.40 | 0.07 | 2.62 | 0.07 | -0.08 | 0.15 |
| px #2 | 0.0633 | 0.0001 | 1.52 | 0.07 | 2.90 | 0.07 | -0.02 | 0.15 |
| px #3 | 0.0350 | 0.0001 | 1.33 | 0.06 | 2.72 | 0.06 | 0.16 | 0.14 |
| px #4 | 0.0223 | 0.0001 | 1.45 | 0.06 | 2.69 | 0.07 | -0.10 | 0.14 |
| px #5 | 0.0233 | 0.0001 | 1.40 | 0.06 | 2.92 | 0.07 | 0.22 | 0.14 |
| px #6 | 0.0330 | 0.0001 | 1.66 | 0.06 | 3.21 | 0.07 | 0.02 | 0.14 |
| px #7 | 0.0438 | 0.0001 | 1.28 | 0.07 | 2.45 | 0.07 | -0.02 | 0.16 |
| px #8 | 0.0448 | 0.0001 | 1.24 | 0.06 | 2.57 | 0.07 | 0.19 | 0.14 |
| px #9 | 0.0425 | 0.0001 | 1.27 | 0.06 | 2.68 | 0.07 | 0.25 | 0.14 |
| px #10 | 0.0776 | 0.0001 | 1.29 | 0.08 | 2.65 | 0.08 | 0.14 | 0.17 |
| px #11 | 0.0261 | 0.0001 | 1.47 | 0.05 | 2.84 | 0.07 | -0.03 | 0.13 |
| average | 0.0407 | 0.0102 | 1.39 | 0.08 | 2.75 | 0.13 | 0.067 | 0.076 |

**Table S7.** Noble gas abundances and isotope ratios for EC 002. Gas was extracted by step heating sequentially to the temperatures indicated. Total isotope ratios are calculated as a weighted average to account for the differential release of gas at different stages of heating.

| T (°C) | Abundances (mol) | | | | | | Isotope ratios | | | | | | | | | |
|---|---|---|---|---|---|---|---|---|---|---|---|---|---|---|---|---|
| | $^3$He (x10$^{18}$) | ±1σ | $^{21}$Ne (x10$^{17}$) | ±1σ | $^{40}$Ar (x10$^{14}$) | ±1σ | $^3$He/$^4$He | ±1σ | $^{20}$Ne/$^{22}$Ne | ±1σ | $^{21}$Ne/$^{22}$Ne | ±1σ | $^{40}$Ar/$^{36}$Ar | ±1σ | $^{38}$Ar/$^{36}$Ar | ±1σ |
| 600 | 3374 | 227 | 67.0 | 3.1 | 14.8 | 0.7 | 0.00871 | 0.00007 | 0.571 | 0.003 | 0.613 | 0.013 | 1026 | 14 | 0.215 | 0.006 |
| 800 | 70160 | 4720 | 89.3 | 4.1 | 1417 | 45 | 0.00741 | 0.00007 | 0.506 | 0.003 | 0.675 | 0.014 | 19701 | 178 | 0.791 | 0.011 |
| 1000 | 862 | 58 | 69.2 | 3.2 | 1139 | 36 | 0.00601 | 0.00007 | 0.624 | 0.004 | 0.880 | 0.019 | 25096 | 236 | 0.997 | 0.014 |
| 1200 | 3907 | 262 | 1217 | 55 | 6390 | 198 | 0.00448 | 0.00003 | 0.807 | 0.003 | 0.870 | 0.018 | 4698 | 33 | 1.218 | 0.016 |
| 1400 | 5.9 | 0.6 | 7.6 | 0.4 | 77.5 | 2.5 | 0.00105 | 0.00011 | 0.613 | 0.014 | 0.767 | 0.019 | 492.5 | 4.3 | 0.395 | 0.005 |
| Total | 78310 | 4730 | 1450 | 56 | 9039 | 207 | 0.00731 | 0.00007 | 0.768 | 0.002 | 0.846 | 0.015 | 5522 | 41 | 1.170 | 0.015 |

**Concentration (mol/g)**

| $^3$He | $^{21}$Ne | $^{40}$Ar |
|---|---|---|
| 9.55 (±0.57) x10$^{-12}$ | 1.77 (±0.07) x10$^{-12}$ | 1.10 (±0.03) x10$^{-8}$ |

**B/ Reflectance spectra of Erg Chech 002 and search for parent bodies.**

**Material and methods**

Reflectance spectra of Erg Chech 002 were measured at Institut de Planétologie et d'Astrophysique de Grenoble (IPAG) using the Shadows instrument (13). We used the standard mode of the instrument (around 7 mm diameter illumination spot) and spectra were measured under nadir illumination and using an observation angle of 30°. One spectrum was obtained for powdered sample and three spectra were obtained on distinct location of a raw slab of the meteorite.

**Results**

All spectra (Fig. S11) showed two strong absorptions around 1 μm (Band I) and 2 μm (Band II) diagnostic of pyroxene. The position of the two bands points toward the presence of clinopyroxene (14). In order to compare the spectra of EC 002 with asteroid observations, we calculated the Band Area Ratio (BAR (15)) and the position of Band I and BII.

**Discussion**

*Space weathering model*- An ensemble of processes referred to as space weathering are known to act at the surface of airless bodies (16). These processes lead to a modification of the optical properties of surface material that may need to be accounted for in the comparison of spectra of meteorites to telescopic observations of small bodies. In the case of silicate-rich asteroids, the model of Hapke (17) enables to simulate numerically these effects, by modeling the incorporation of small opaque grains to the measured spectra. The Hapke model (17) was applied to the spectra of EC 002 powders, with different magnitude of space weathering, corresponding to different amounts of opaque phases incorporated in the silicate grains. The spectra resulting from these simulations are shown in figure S12 (top). These simulations reveal a progressive darkening and reddening of the spectra together with a decrease of the silicate absorptions (Figure S12 top).

*Spectral effects of adding olivine xenocrysts*- Because some samples of EC 002 contain olivine xenocrysts, we investigated numerically the effect of adding olivine crystals to the measured spectra of EC 002. For that, we used the model of Hapke (19). A reflectance spectra of San Carlos olivine of known grain size was used, and numerically mixed to the spectra of EC 002. Mixture were performed in single-scattering albedo space, in order to simulate the spectrum an intimate mixture. Results are shown in figure S12 bottom, in the cases of adding 5,10 and 20 vol. % of olivine to our measured EC 002 spectrum.

*Spectral effects of adding fragments of a residual mantle-* In order to model the effect of mixing EC 002 with fragments of excavated mantle, we proceeded as follow. The mantle signature was calculated using reflectance spectra of olivine (Fo90, RELAB database) mixed in equal proportion with orthopyroxene (En80, RELAB database). The mixture of olivine and pyroxene was computed in single-scattering albedo space to model an intimated mixture. This calculated mantle spectrum was then mixed with the reflectance spectrum measured for EC 002 powder using a linear mixing model (Fig. S13).

*Comparison to taxonomic endmembers*- In figure S12 bottom, the spectra of EC 002 are compared to the spectra of asteroid taxonomic endmembers with strong pyroxene silicate signatures, namely O- and V- type asteroids (Fig S11 bottom). None of the taxonomic endmembers identified by DeMeo et al. (18) appears to correspond to the spectra measured for EC 002 (Fig S11 bottom). To push the comparison further, the BAR vs Band I position diagram can be used following Gaffey et al. (15). This analysis confirms the lack of spectral matches among taxonomic endmembers (Fig. S13). It also reveals that the spectra of EC 002 is distinct from that of HED meteorites, with a higher position of Band I, and a lower Band Area Ratio.

The models described earlier enables to draw the potential evolution of EC 002 in this diagram as a response of olivine addition or space-weathering. Space-weathering is not able to move the location of EC 002 toward any know taxonomic endmember. In the case of addition of olivine, the position of EC 002 moves close to the rare O-type endmember. However, the amount of olivine needed is quite high (20 %) when comparing to the typical amount of xenolitic olivine, when present. Note that O-type asteroids are extremely rare and include at the moment only a few objects.

***Comparison to Sloan Digital Sky Survey (SDSS)-*** Color surveys while less powerful in term of composition, offer the capability to observe several orders of magnitudes more objects than spectroscopy. Because of the strong absorption feature at 1-µm in the spectra of EC 002, color survey including an infrared filter enable to search for possible parent bodies within a much higher number of asteroids. For that, we followed the procedure describe in DeMeo and Carry (20) that was used to map the distribution of asteroids spectral types across the Solar System (21). We used data from the SDDS, selected following the same analysis as in DeMeo and Carry (20). Colors of EC 002 and different classes of meteorites were also computed and plot in the z-i vs g'r'i' slope diagram. Spectra of HED meteorites and ordinary chondrites were extracted through the RELAB database (22). This plot (Fig. S14) reveals that the colors of EC 002 lie outside of the asteroid endmember boundaries used in DeMeo and Carry (20) and that the vast majority of asteroids do not match the composition of Erg Chech. Space weathering can explain the mismatch of HED meteorites and V-type asteroids, but in the case of EC 002, space weathering models do not "push" the colors of EC 002 in one of the asteroid endmembers. As concluded from spectroscopy, color analysis furthers strength the lack of EC 002-like asteroid within the Solar System.

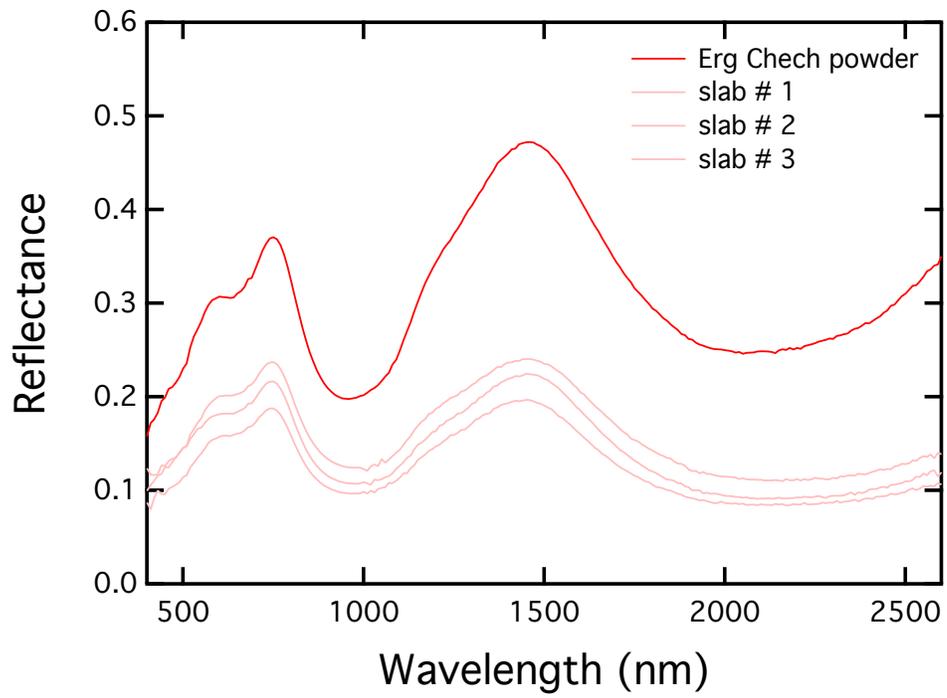
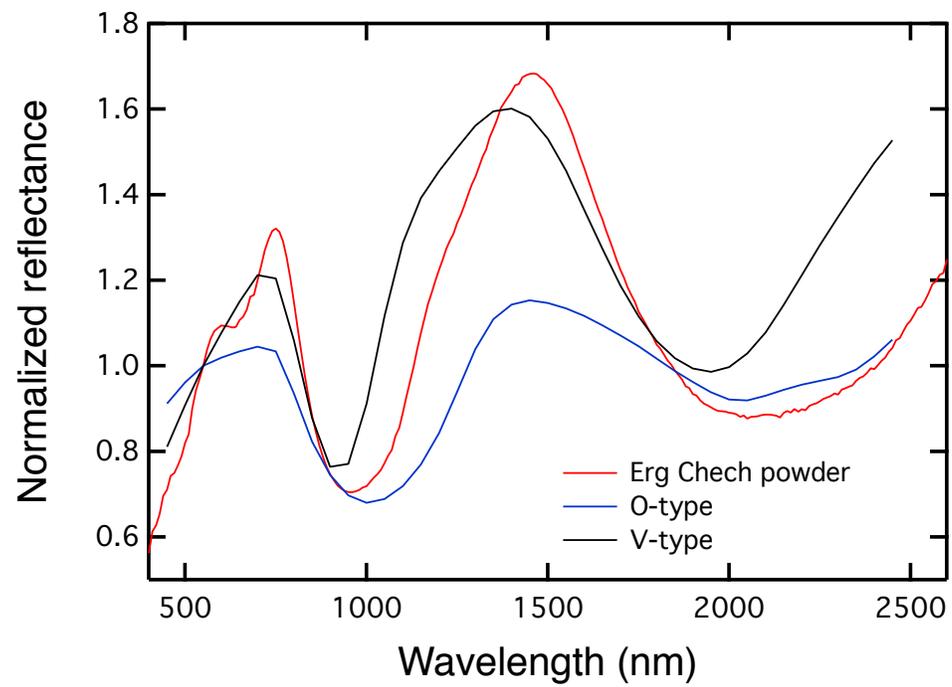

**Fig. S11**. Top) Reflectance spectra of EC 002 obtained for powdered sample, and 3 distinct locations of a raw piece of sample. Bottom) Reflectance spectra of EC 002 compared to V- and O- type endmembers (18)

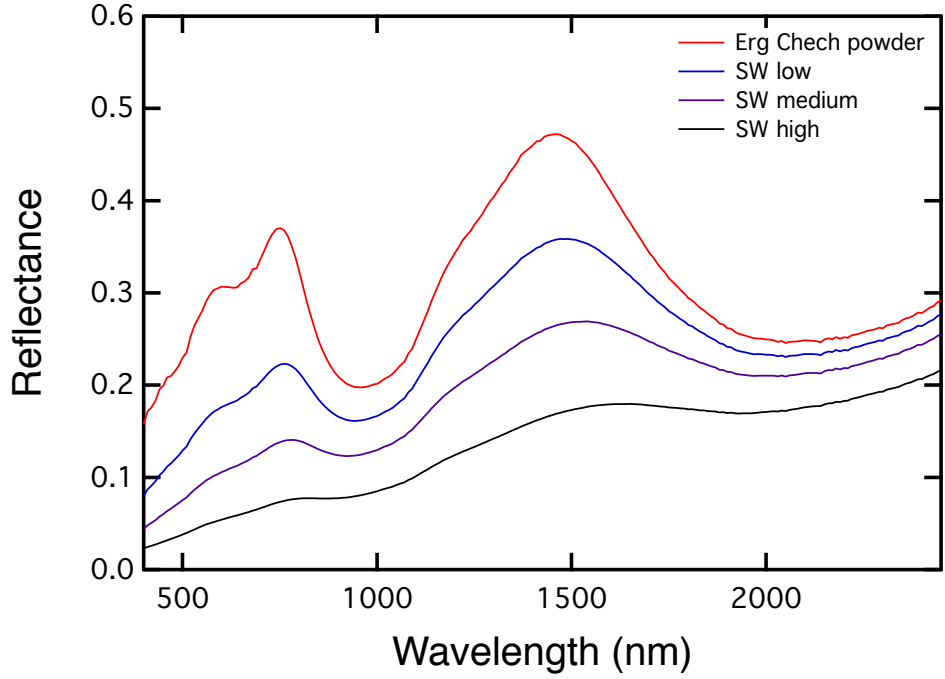

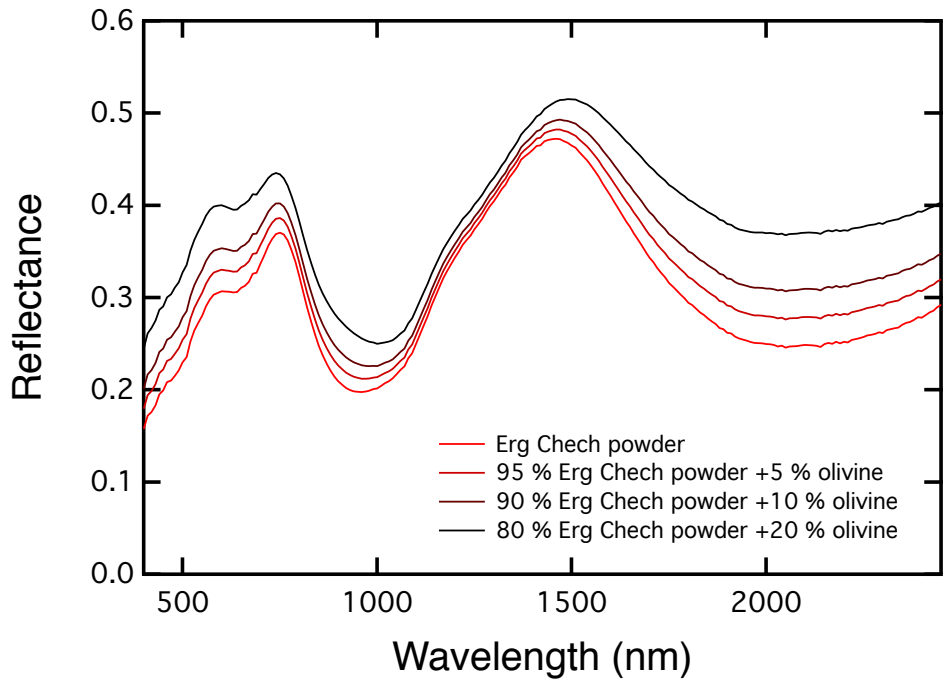

Fig. S12: Modeled spectra of EC 002 considering a space-weathering model (top) and mixture with olivine crystals (bottom)

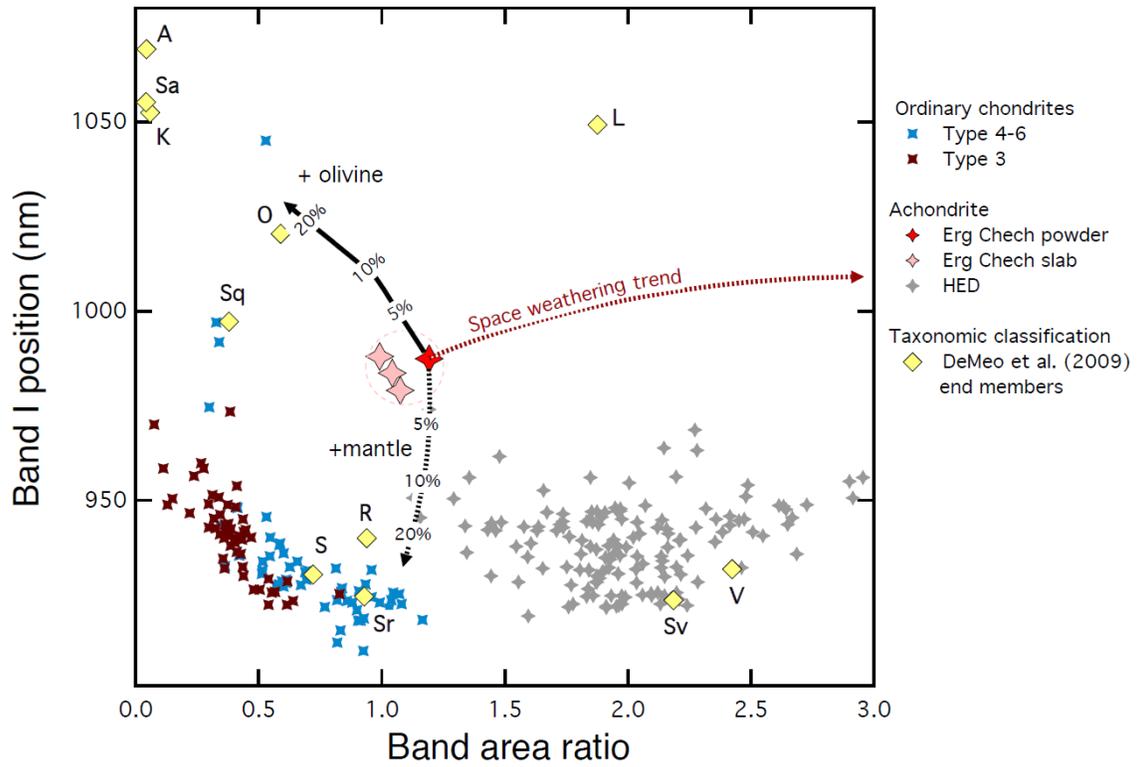

Fig. S13. BII/BI area ratio as a function of the Band I position. Parameters derived for EC 002 are compared to those of HED meteorites and ordinary chondrites (RELAB database). The positions of DeMeo et al. (20) asteroid endmembers are also shown, and the trends that were calculated based on mixtures with olivine, mantle, and using a space weathering model (17).

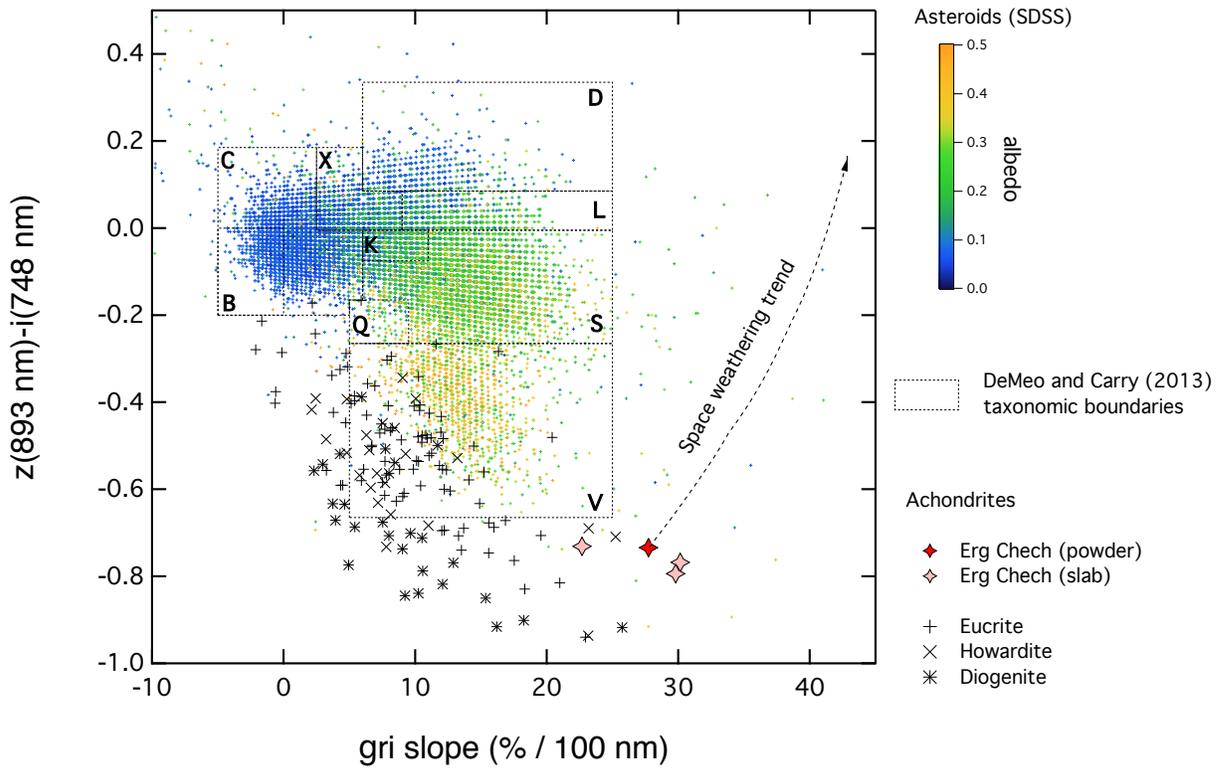

Fig. S14. Color diagram used by DeMeo and Carry (20) for classifying asteroid from the SDSS (asteroid are color coded here according to their wise albedo (22), compared to Erg Chech and HED meteorite. This graph testifies of the scarcity of Erg Chech -like objects, even if possible effects of space weathering are taken into account.

**SI References**

1. J. Cotten et al., Origin of anomalous Rare-Earth Element and Yttrium enrichments in subaerially exposed basalts - evidence from French-Polynesia. *Chem. Geol.* 119, 115–138 (1995).

2. J.-A. Barrat, et al., Geochemistry of CI chondrites: Major and trace elements, and Cu and Zn isotopes. *Geochim Cosmochim Acta* 83, 79–92 (2012).

3. J.A. Barrat, et al., Evidence from Tm anomalies for non-CI refractory lithophile element proportions in terrestrial planets and achondrites. *Geochim. Cosmochim. Acta* 176, 1-17 (2016).

4. E. D. Young, A. Galy, The Isotope Geochemistry and Cosmochemistry of Magnesium. *Rev. Mineral. Geochemistry* 55, 197--230. (2004).

5. M. W. Broadley et al., Noble gas variations in ureilites and their implications for ureilite parent body formation. *Geochimica et Cosmochimica Acta* 270, 325–337 (2020).

6. I. Leya et al., The production of cosmogenic nuclides by galactic cosmic-ray particles for 2π exposure geometries. *Meteoritics & Planetary Science* 36, 1547–1561 (2001).

7. S. Kelley, K-Ar and Ar-Ar dating. *Reviews in Mineralogy and Geochemistry* 47, 785–818 (2002).

8. U. Ott, Noble Gases in Meteorites – Trapped Components. *Reviews in Mineralogy and Geochemistry* 47, 71–100 (2002).

9. J.M.D. Day, et al., Early formation of evolved asteroidal crust. *Nature* 457(7226), 179-182 (2009).

10. A. Bischoff, et al., Trachyandesitic volcanism in the early Solar System. *Proc. Natl. Acad. Sci. U.S.A.* 111, 12689-12692 (2014).

11. P. Srinivasan, *et al.*, Silica-rich volcanism in the early solar system dated at 4.565 Ga. *Nature Commun.* 9**,** 3036 (2018).

12. M. Collinet, T. L. Grove, Widespread production of silica- and alkali-rich melts at the onset of planetesimal melting. *Geochim. Cosmochim. Acta* 277, 334-357 (2020).

13. S. Potin, *et al.*, SHADOWS: a spectro-gonio radiometer for bidirectional reflectance studies of dark meteorites and terrestrial analogs: design, calibrations, and performances on challenging surfaces. *Applied Optics* 57*, issue 28, p. 8279* 57, 8279 (2018).
14. J. B. Adams, Visible and near-infrared diffuse reflectance spectra of pyroxenes as applied to remote sensing of solid objects in the solar system. *Journal of Geophysical Research* 79, 4829 (1974).
15. M. J. Gaffey et al., Mineralogy of Asteroids. *Asteroids III*, 183–204 (2002).
16. C. M. Pieters, S. K. Noble, Space weathering on airless bodies. *Journal of Geophysical Research (Planets)* 121, 1865–1884 (2016).
17. B. Hapke, Space weathering from Mercury to the asteroid belt. *Journal of Geophysical Research* 106, 10039 (2001).
18. F. E. DeMeo, R. P. Binzel, S. M. Slivan, S. J. Bus, An extension of the Bus asteroid taxonomy into the near-infrared. *Icarus* 202, 160–180 (2009).
19. B. Hapke, *Theory of reflectance and emittance spectroscopy*, Cambridge University Press 469 pages (1993)
20. F. E. DeMeo, B. Carry, The taxonomic distribution of asteroids from multi-filter all-sky photometric surveys. *Icarus* 226, 723-741 (2013).
21. F. E. DeMeo, B. Carry, Solar System evolution from compositional mapping of the asteroid belt. *Nature* 505, 629–634 (2014).
22. J. R. Masiero, *et al.*, Main Belt Asteroids with WISE/NEOWISE. I. Preliminary Albedos and Diameters. *The Astrophysical Journal* 741, 68 (2011).